\documentclass[aps,prc,a4paper,nofootinbib,showpacs,showkeys,
preprintnumbers,superscriptaddress,twocolumn]{revtex4}

\usepackage{amsmath}
\usepackage{amssymb}
\usepackage{bm}
\usepackage{graphicx}
\usepackage{color}
\usepackage[english]{babel}

\newcommand{\er}{~\eqref}

\begin{document}

\title{Quark Matter in a Parallel Electric and Magnetic Field Background:\\
Chiral Phase Transition and Equilibration of Chiral Density}

\author{M.~Ruggieri}\email{marco.ruggieri@ucas.ac.cn}
\affiliation{College of Physics, University of Chinese Academy of Sciences, 
Yuquanlu 19A, Beijing 100049, China}

\author{G.~X.~Peng}\email{gxpeng@ucas.ac.cn}
\affiliation{%
College of Physics, University of Chinese Academy of Sciences, 
Yuquanlu 19A, Beijing 100049, China}
\affiliation{Theoretical Physics Center for Science Facilities, Institute of High Energy Physics, Beijing 100049, China}


\begin{abstract}
In this article we study spontaneous chiral symmetry breaking for quark matter
in the background of static and homogeneous parallel electric field $\bm E$ and magnetic field $\bm B$.
We use a Nambu-Jona-Lasinio model with a local kernel interaction to compute the relevant quantities
to describe chiral symmetry breaking at finite temperature for a wide range
of $E$ and $B$. We study the effect of this background 
on inverse catalysis of chiral symmetry breaking for $E$ and $B$ of the
same order of magnitude. We then focus on
the effect of equilibration of chiral density, $n_5$, produced dynamically by axial anomaly on the critical temperature.
The equilibration of $n_5$, a consequence of chirality flipping processes in the thermal bath, 
allows for the introduction of the chiral chemical potential, $\mu_5$,
which is computed self-consistently as a function of temperature and field strength
by coupling the number equation to the gap equation, and solving the two within an expansion in
$E/T^2$, $B/T^2$ and $\mu_5^2/T^2$.
We find that even if chirality is produced and equilibrates within a relaxation time $\tau_M$, 
it does not change drastically the thermodynamics, with particular reference to the inverse catalysis
induced by the external fields, as long as the average $\mu_5$ at equilibrium is not too large.

\end{abstract}


\maketitle

\section{Introduction}

There has been recently an increasing interest for study of systems with 
a finite chiral density, namely $n_5\equiv n_R-n_L \neq 0$.
Such chirality imbalance can be obtained dynamically because of the Adler-Bell-Jackiw 
anomaly~\cite{Adler:1969gk,Bell:1969ts} when fermions interact with nontrivial
gauge field configurations characterized by a topological index named the winding number, $Q_W$.  
In the context of Quantum Chromodynamics (QCD)  such nontrivial gauge field configurations 
at finite temperature in Minkowski space are
named sphalerons, whose production rate has been estimated to be quite large~\cite{Moore:2000ara,Moore:2010jd} .
The large number of sphaleron transitions in high temperature
suggests the possibility that net chirality might be abundant (locally) in the quark-gluon plasma phase of QCD;
when one couples this thermal QCD bath with an external strong magnetic field, $\bm B$,
produced in the early stages of heavy ion collisions,
the coexistence of $n_5\neq 0$ and $\bm B\neq 0$
might lead to a charge separation phenonemon named the Chiral Magnetic Effect
(CME)~\cite{Kharzeev:2007jp,Fukushima:2008xe} which has been observed experimentally
in zirconium pentatelluride~\cite{Li:2014bha}. Beside CME other interesting effects related to
anomaly and chiral imbalance can be found in~\cite{Son:2009tf,Banerjee:2008th,Landsteiner:2011cp,Son:2004tq,
Metlitski:2005pr,Kharzeev:2010gd,Chernodub:2015gxa,Chernodub:2015wxa,Chernodub:2013kya,Braguta:2013loa,
Sadofyev:2010pr,Sadofyev:2010is,Khaidukov:2013sja,Kirilin:2013fqa,Avdoshkin:2014gpa}.

In order to describe systems with finite chirality in thermodynamical equilibrium,
it is customary to introduce a chiral chemical potential, $\mu_5$,
conjugated to the $n_5$~\cite{Ruggieri:2016cbq,Ruggieri:2016ejz,Frasca:2016rsi,
Gatto:2011wc,Fukushima:2010fe,Chernodub:2011fr,Ruggieri:2011xc,Yu:2015hym,Yu:2014xoa,
Braguta:2015owi,Braguta:2015zta,Braguta:2016aov,Hanada:2011jb,Xu:2015vna}. 
The chiral chemical potential describes a system in which chiral density is in thermodynamical equilibrium;
however because of anomaly as well as of chirality changing processes due to finite quark condensate,
$n_5$ is not a strictly conserved quantity hence the meaning of $\mu_5$ is not so clear; 
however naming $\tau_M$ the typical time scale in which
chirality changing processes take place, one might assume
that $\mu_5\neq0$ describes a system in thermodynamical equilibrium with a fixed value 
of $n_5$ on a time scale much larger than $\tau_M$, the latter representing the time scale
needed for $n_5$ to equilibrate.

In this article we study chiral phase transition and chiral density production in the context of
quark matter in a background static and homogeneous parallel electric, $\bm E$, and magnetic, $\bm B$, fields.
One of our goal is to investigate the effect of the background fields 
on chiral symmetry breaking at zero temperature,
and on the critical temperature for chiral symmetry restoration, $T_c$.
This part of the study embraces previous studies about chiral symmetry breaking/restoration in the background
of external fields~\cite{Babansky:1997zh,Klevansky:1989vi,Suganuma:1990nn,
D'Elia:2012zw,Cao:2015cka,Gusynin:1994xp,Gusynin:1994re,Krive:1992xh,Klimenko:1992ch,Klimenko:1991he}, 
completing them by adding the computation of the critical temperature
versus the strengh of $E$ and $B$. We find that the effect of the electric field is to lower the critical temperature,
in agreement with the scenario of inverse catalysis depicted in \cite{Klevansky:1989vi,Suganuma:1990nn,Cao:2015dya}
where however the zero temperature case has only been considered;
the inverse catalysis scenario does not change considerably when the magnetic field is added, 
as long as the magnetic field is not very large compared to the electric one. 
This finding is in agreement with a previous study at zero temperature~\cite{Babansky:1997zh}
where the role of the second electromagnetic invariant, $\bm E\cdot\bm B$, has been recognized
as inhibitor of chiral symmetry breaking.

We are also interested to study the effect of chiral density 
on the thermodynamics of the system.
The model studied here has the advantage that a chiral density is obtained dynamically without the need
to introduce, {\it a priori}, a chiral chemical potential. As a matter of fact 
chirality can be produced combining $\bm E$ which produces pairs via the Schwinger mechanism, 
and $\bm B$ which alignes particles spin along its direction. 
The mechanism producing chirality is very simple: we assume for sake of simplicity 
a very large $\bm B$, so that only the lowest Landau level (LLL) is occupied;
moreover we assume the system made only of one flavor of quarks, namely $u$ quarks,
and we focus on a single $u\bar u$ created by the Schwinger effect.
The $u$ quark must have its spin aligned along $\bm B$
because it sits in the LLL, and its momentum
will be initially rather parallel or antiparallel to $\bm B$, so the initial helicity can be either positive or negative.
On the other hand the effect of $\bm E\parallel \bm B$ is to accelerate 
$u$ along the direction of $\bm B$ so after some time $u$ quark will have positive helicity. An analogous discussion
can be done for the $\bar u$. 
Therefore as a consequence of the Schwinger effect, LLL and $\bm E\parallel \bm B$ each time a pair is created,
there is an increase of a factor two of the net chiral density of the system.

The dynamical evolution of $n_5$ produced by this mechanism can be computed explicitly~\cite{Warringa:2012bq} 
and it has been shown to be the one expected from the 
Adler-Bell-Jackiw anomaly: this is not surprising because $\bm E \cdot \bm B \neq 0$
meaning that axial current is not conserved at the quantum level and $n_5$ should evolve according to the 
anomaly equation. If $n_5$ evolution was governed only by the anomaly, however, there would be no chance
for reaching a thermodynamical equilibrium because $n_5$ would grow indefinitely (assuming the fields as external
fields and neglecting any backreaction from the fermion currents). 
But in the thermal bath there are also chirality flipping processes related to the existence of the chiral condensate
as well as of the finite current quark mass: we introduce a relaxation time for chirality, namely $\tau_M$, 
giving the time scale necessary for the equilibration of $n_5$. Then it is possible to show that
for times $t\gg\tau_M$ chiral density equilibrates to $n_5^{\mathrm{eq}}$, 
the actual value depending on quark electric charge,  fields magnitude and temperature.

Because $n_5$ equilibrates it is possible to introduce the chiral chemical potential, $\mu_5$, conjugated to $n_5^{\mathrm{eq}}$ at equilibrium.
Differently from previous calculations with chirality imbalance, in the present study 
we compute the value of $\mu_5$
self-consistently by coupling the gap equation to the number equation, 
even if we limit ourselves to the approximation of small fields and small $\mu_5$,
namely working at the leading order in $\mu_5/T$ and $E/T^2$, $B/T^2$.
As a consequence, $\mu_5$ will depend on temperature as well as on external fields, and on the relaxation time.
We focus on the effects of the external fields on the chiral phase transition, with emphasis on the
role of chirality production in the critical region. Because of the small fields approximation involved in the
solution of the gap as well as the number equations, we are aware that our picture about thermodynamics
might change in case of large fields.

In this study we compute the effect of the dynamically produced $n_5$ on $T_c$.
As mentioned above, the $\bm E \cdot \bm B$ term tends to lower the critical
temperature; on the other hand the chiral chemical
potential has the effect to increase $T_c$
\cite{Ruggieri:2016cbq,Ruggieri:2016ejz,Frasca:2016rsi,
Yu:2015hym,Yu:2014xoa,Braguta:2015owi,Braguta:2015zta,Braguta:2016aov}. 
Therefore, it is interesting to compute the response of $T_c$ to the simultaneous presence of
$\mu_5$ and fields, to check if the inverse catalysis scenario obtained at $\mu_5 = 0$ still
persists at $\mu_5 \neq 0$. 
We can anticipate our results, namely that chiral density does not affect drastically the thermodynamics 
at the phase transition, confirming the inverse catalysis induced by the fields,
as long as the average chiral chemical potential in the crossover region turns out to be small with respect to temperature.
In Section V we present a detailed study of this effect, showing concrete numbers and
among other things how changing  the field strenghts and/or the relaxation time magnitude affects the inverse catalysis. 
In fact we have found and report about situations in which we can measure a net effect of the chiral chemical potential
on the constituent quark mass and on critical temperature, even if we take these results with a grain of salt
as the value of $\mu_5$ at equilibrium turns out to be of the order of the critical temperature, hence
potentially validating our quantitative predictions.

The relaxation time for chirality adds the greatest theoretical uncertainty to our calculations: 
in absence of a specific calculation of $\tau_M$ it is possible to give only a rough estimate based on 
dimensional analysis as well as on physical reasons;
we chose $\tau_M \propto 1/M_q$ where $M_q$ is the constituent quark mass which is computed self-consistently
within the model: it depends on temperature and fields, and by construction it brings informations about the
chiral condensate at zero as well as finite temperature. 
Because of this uncertainty on $\tau_M$ we feel it is not so important, in this explorative study, 
to present the most complete calculation possible taking into account the full propagators with the 
full $\mu_5$ dependence: we suspect in fact that even within the most accurate calculation possible,
the new effects of the chiral density on the phase transition might be cancelled by changing $\tau_M$
which still would remain unknown. We therefore prefer to limit ourselves to a simple weak fields
and small $\mu_5$ approximation to explore the effects the chiral density will have on the phase diagram,
leaving a more complete calculation to a future study.   

The plan of the article is as follows. In Section II we briefly review the model we use for our calculations.
In Section III we present few selected results at zero temperature which show the interplay between the electric
and magnetic fields on chiral symmetry breaking. In Section IV we discuss some result at finite temperature,
with emphasis on the chiral phase transition without taking into account chirality production.
In Section V we compute chirality at equilibrium and the related chiral chemical potential, and study the effect
of this chirality on the critical temperature. Finally in Section VI we draw our Conclusions.

\section{The model}
In this article we are interested to study quark matter in a background of an electric-magnetic fluc tube made of
parallel electric, $\bm E$, and magnetic, $\bm B$, fields. 
We assume the fields are constant in time and homogeneous in space; moreover we assume they develop along the 
$z-$direction. 
In this Section we describe the model we use for our calculations. 
More specifically, we use a Nambu-Jona-Lasinio (NJL) model~\cite{Nambu:1961tp,Nambu:1961fr}
(see~\cite{Klevansky:1992qe,Hatsuda:1994pi} for reviews) with a local interaction kernel, in which we introduce
the coupling of quarks with the external alectric and magnetic fields. The set up of the gap equation has been presented
in great detail in~\cite{Cao:2015dya} which we follow, therefore we will skip all the technical details and report here only
the few equations we need to specify the interactions used in the calculations.
The Euclidean lagrangian density is given by
\begin{equation}
{\cal L}=\bar\psi
\left(
i D\!\!\!\!/ - m_0 
\right)\psi + G\left[(\bar\psi\psi)^2 + (\bar\psi i \gamma_5 \bm\tau\psi)^2 \right],
\end{equation}
with $\psi$ being a quark field with Dirac, color and flavor indices, $m_0$ is the current quark mass
and $\bm\tau$ denotes a vector of Pauli matrices on flavor space. 
The interaction with the background fields is embedded in the covariant derivative 
$D\!\!\!\!/ = (\partial_\mu -i A_\mu \hat q)\gamma_\mu$, where $\gamma_\mu$ denotes
the set of Euclidean Dirac matrices and $\hat q$ is the quark electric charge matrix in flavor space.
In this work we use the gauge $A_\mu = (i E z,0,-B x,0)$. 

Introducing the auxiliary field $\sigma= -2G\bar\psi\psi$ and using a mean field approximation,
the thermodynamic potential can be written as
\begin{equation}
\Omega = \frac{(M_q - m_0)^2}{4G} - \frac{1}{\beta V}\text{Tr}\log\beta(i D\!\!\!\!/ - M_q),
\end{equation}
where the constituent quark mass is $M_q = m_0 -2G\langle\bar\psi\psi\rangle$, $\beta=1/T$ and
$\beta V$ corresponds to the Euclidean quantization volume. The constituent quark mass differs from $m_0$
because of spontaneous chiral symmetry breaking, the latter being related to a nonvanishing
chiral condensate, $\langle\bar\psi\psi\rangle\neq0$. Even if it would be more appropriate to discuss
chiral symmetry restoration via the quark condensate, because it has its counterpart in QCD, in this article
we will refer to $M_q$ for simplicity, keeping in mind that whenever we discuss about the chiral phase transition
in terms of $M_q$ the decrease of the latter is related to the decreasing chiral condensate.

In this model the main task is to compute self-consistently $M_q$ at finite temperature and in presence of the
external fields. This is achieved by requiring the physical value of $M_q$ minimizes the thermodynamic potential,
and this in turn implies that $M_q$ satisfies the gap equation, $\partial\Omega/\partial M_q = 0$, 
namely
\begin{equation}
\frac{M_q -m_0}{2G} - \frac{1}{\beta V}\text{Tr} {\cal S}(x,x^\prime)=0,
\end{equation}
where ${\cal S}(x,x^\prime)$ corresponds to the full fermion propagator in the electric and magnetic field background.
The computation of the propagator has been already given in detail in~\cite{Cao:2015dya} therefore here we merely quote
the final result for the gap equation, that is
\begin{eqnarray}
\frac{M_q-m_0}{2G}&=&M_q \frac{N_c}{4\pi^2}\sum_f \int_0^\infty  \frac{ds}{s^2} e^{-M_q^2 s}{\cal F}(s)
\nonumber\\
&&+M_q \frac{N_c N_f}{4\pi^2}\int_{1/\Lambda^2}^\infty\frac{ds}{s^2} e^{-M_q^2 s},
\label{eq:gkk}
\end{eqnarray}
where we have introduced the functions
\begin{equation}
{\cal F}(s) = \theta_3
\left(
\frac{\pi}{2},e^{-|{\cal A}|}
\right)\frac{q_f eB s}{\tanh(q_f e B s)}\frac{q_f eE s}{\tan(q_f e E s)}-1
\label{eq:Fs}
\end{equation}
with $\theta_3(x,z)$ being the third elliptic theta function, and
\begin{equation}
{\cal A}(s) = \frac{q_f e E}{4T^2\tan(q_f e E s)}.
\end{equation}
In Eq.\er{eq:gkk} we have added and subtracted the zero field contribution on the right hand side which 
is the only one to diverge, and we have regularized it by cutting the integration at $s=1/\Lambda^2$; 
on the other hand we have not added
a cutoff on the field dependent part as it is not divergent. 
For the parameters choice we use the standard parameter set for a proper time
regularization~\cite{Klevansky:1992qe}, namely $\Lambda = 1086$ MeV and $G=3.78/\Lambda^2$.

The presence of the $1/\tan(q_f e E s)$ in Eq.\er{eq:Fs} implies the existence of an infinite set of poles
on the integration in $s$ in Eq.\er{eq:gkk}; these poles appear in $\Omega$ as well.
Following the original treatment by Schwinger~\cite{Schwinger:1951nm} 
these poles are moved to the complex plane by adding a small imaginary part
which allows to perform the $s-$integration in principal value; this leads to an imaginary part
of the free energy, which is a sign of the vacuum instability induced by the static 
electric field~\cite{Heisenberg:1935qt,Schwinger:1951nm} and leads to particle pair creation. 
We will consider the effect of this vacuum instability in Section~\ref{eq:cd} because it can be directly
connected to chiral density production in case of parallel $\bm E$ and $\bm B$.

\section{Results at zero temperature}

\begin{figure}[t!]
\begin{center}
\includegraphics[width=7cm]{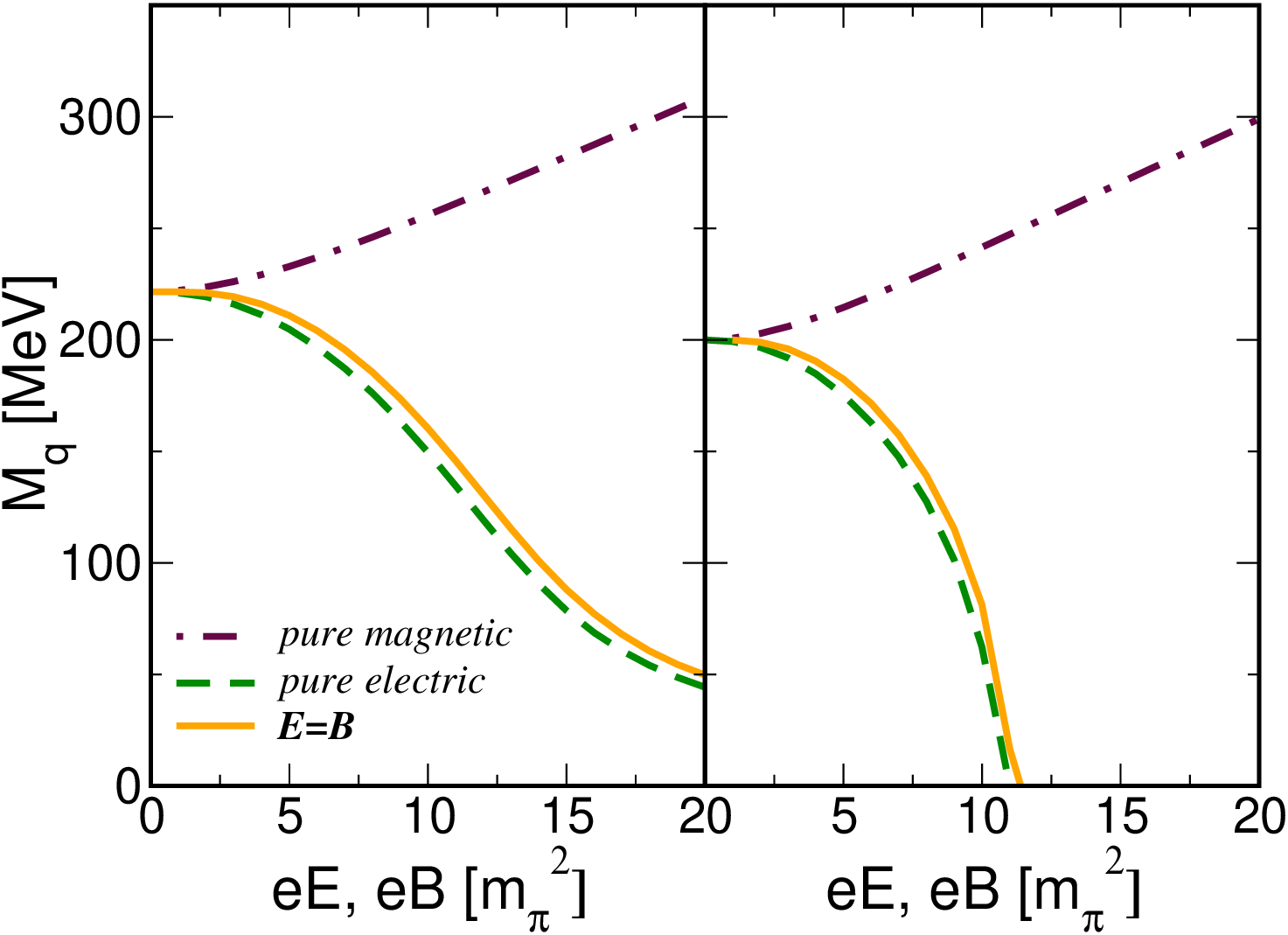}
\end{center}
\caption{\label{Fig:1} Dynamical quark mass with electric and/or magnetic field
strength at zero temperature. Maroon dot-dashed line corresponds to the case of a pure magnetic field;
green dashed line to a pure electric field; finally solid orange line corresponds to the case $E = B$.
Left panel corresponds to $m_0=5.49$ MeV, while right panel corresponds to the $m_0=0$.}
\end{figure}

In this Section we present few results at zero temperature.
In Fig.~\ref{Fig:1} we plot the constituent quark mass as a function of the external field strength at $T=0$
for several cases: maroon dot-dashed line corresponds to the case of a pure magnetic field;
green dashed line to a pure electric field; finally solid orange line corresponds to the case $E = B$.
Left panel corresponds to $m_0=5.49$ MeV which is the value of the current quark mass necessary to have
$m_\pi=139$ MeV; right panel corresponds to the chiral limit $m_0=0$.
For the case of a pure magnetic field we find the magnetic catalysis of chiral symmetry breaking;
on the other hand the electric field has the opposite effect leading to an inverse magnetic catalysis~\cite{Klevansky:1992qe}.
In this pure electric field case there exists a critical electric field at which chiral symmetry is restored
in the chiral limit: we find the transition to be of the second order. In the case of massive quarks the phase transition
is changed into a smooth crossover characterized by a smooth but net change in the slope of the condensate,
resulting in smaller value of the condensate itself,
as it happens for the chiral phase transition at finite temperature. In this case it is not possible to define in a rigorous way
a critical field, but it is still possible to identify a range of electric fields in which $M_q$ has its highest change with $E$, 
and identify this range with the pseudo-critical region.

\begin{figure}[t!]
\begin{center}
\includegraphics[width=7cm]{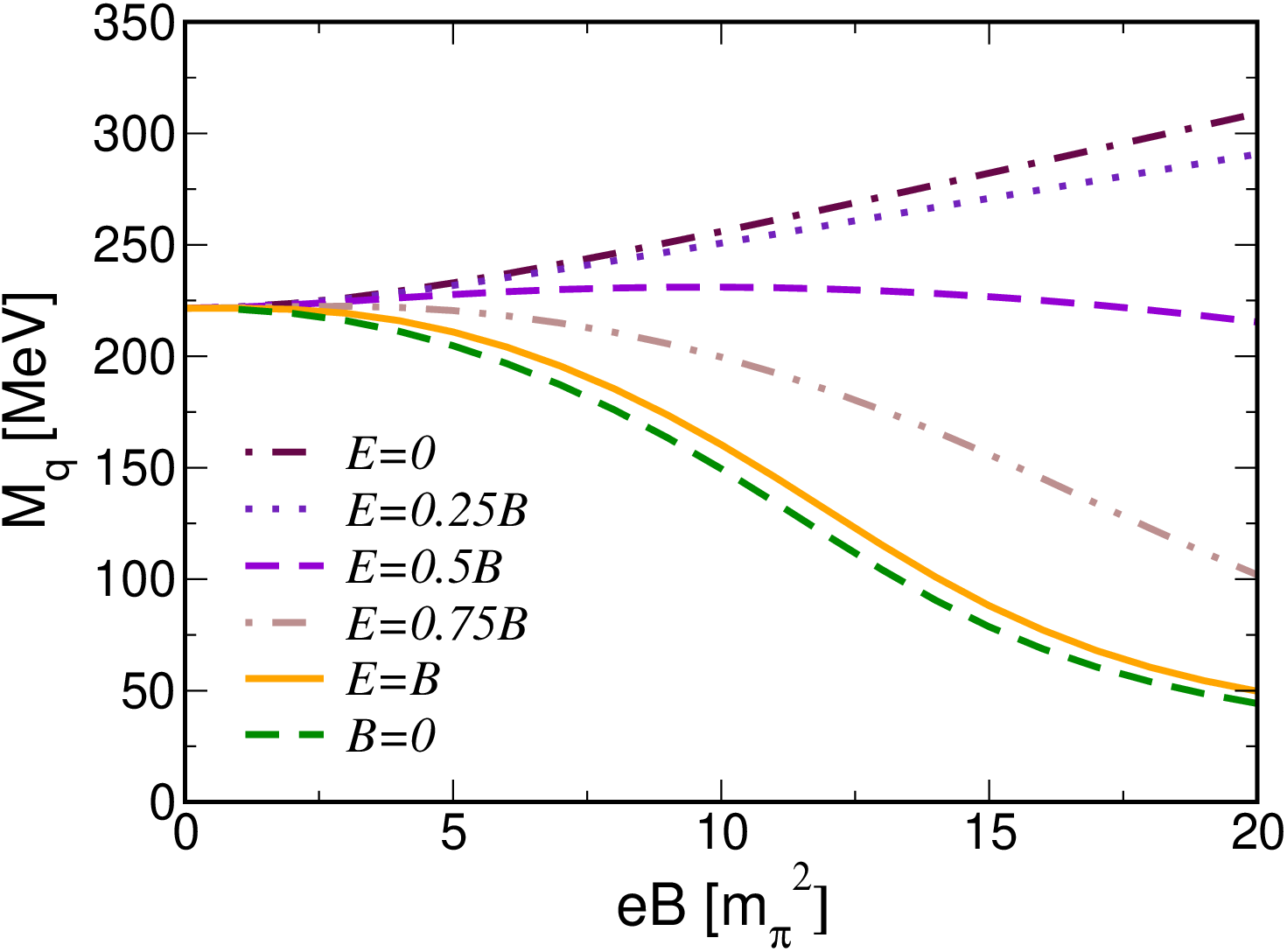}\\
\includegraphics[width=7cm]{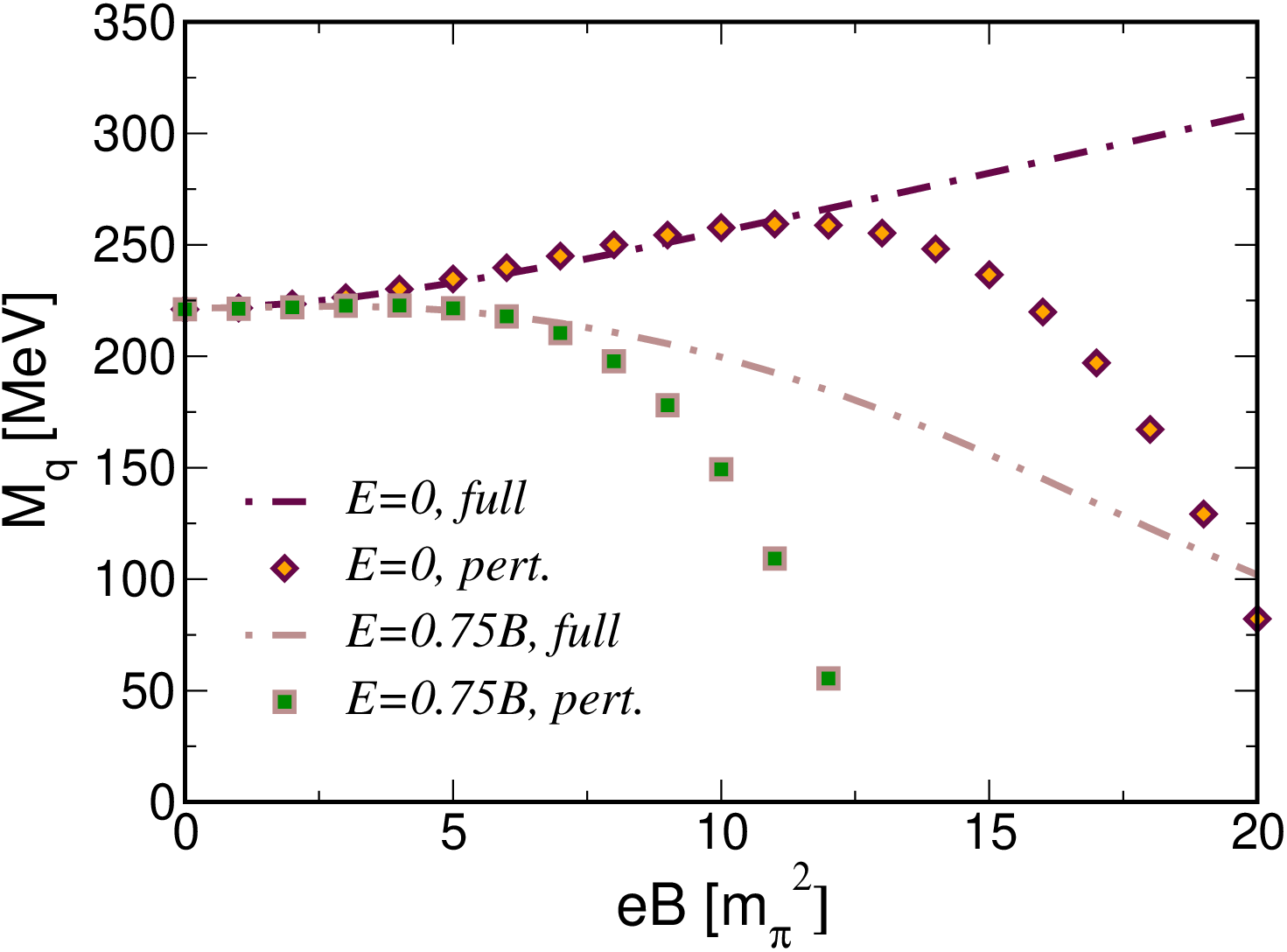}
\end{center}
\caption{\label{Fig:1b} {\it Upper panel.} Dynamical quark mass versus magnetic field
strength at zero temperature, for several values of the background electric field. 
Maroon dot-dashed line corresponds to the case of a pure magnetic field;
indigo dotted line to $E = 0.25 B$; dashed magenta line to $E = 0.5 B$;
dot-dot-dashed brown line to $E = 0.75B$; finally orange line to $E = B$.
For comparison we have also shown data for $B=0$ borrowed from Fig.\ref{Fig:1};
in this case on $x-$axis we show $eE$ in units of $m_\pi^2$.
{\it Lower panel.} Comparison of the perturbative solution Eq.\er{eq:l2} with the full one.}
\end{figure}

It is interesting to study what happens when $E$ and $B$ act together: naturally one would expect
a competition among the effects of the magnetic (catalysis) and electric (inverse catalysis) fields.
In Fig.~\ref{Fig:1} we have shown the case $E = B$ in which it is clear that, regardless we work in the
chiral or in the physical current quark mass limit, the magnetic field has some catalysis effect increasing the value
of $M_q$ (i.e. chiral condensate) and shifts the critical (or pseudo-critical) value of the electric field
slightly upwards compared to the case $B=0$. In Fig.~\ref{Fig:1b} we show $M_q$
as a function of $eB$ for several choices of $E$, starting from $E=0$ up to $E=B$.
Already for $E = 0.5 B$ we find a sign of competition among direct and inverse catalysis,
which manifests in a non-monotonic behaviour of $M_q$ versus $eB$.
We can also read the results of Fig.~\ref{Fig:1b} in the opposite way: given a background of an
electric field $ E$, even introducing a magnetic field of the same magnitude of $ E$ does not result
in a considerable change of spontaneous chiral symmetry breaking, compare green dashed
and orange solid lines in Fig.~\ref{Fig:1}; for $B$ as large as $\approx1.3E$
we find that qualitatively the behaviour of the chiral condensate versus field strength is like the one at $B=0$,
even if for very small values of the field strength we still find the mass increases;
the net effect of the magnetic field is to shift the critical value of the electric field to larger values because of catalysis.
In order to measure a catalysis effect one has to introduce a larger magnetic field, for example $B\approx2E$
in Fig.~\ref{Fig:1b}. The inverse catalysis effect induced by the electric field and the
second electromagnetic invariant, $\bm E\cdot \bm B$, are in agreement with previous studies
at zero temperature~\cite{Babansky:1997zh,Klevansky:1989vi,Suganuma:1990nn,Klimenko:1992ch,Klimenko:1991he}.

The behaviour of $M_q$ for small values of the fields can be easily understood quantitatively by the gap equation
at $T=0$ and $m_0=0$.
We can find an analytical solution for the gap equation\er{eq:gkk} for small fields by writing
$M_q=M_0+\delta m$ where $M_0$ corresponds to the solution of the gap equation for $E = B=0$. 
Moreover for small values of the fields we can keep only the order $O(M_0)$ in the field dependent term in Eq.\er{eq:gkk}.
Taking into account that $M_0$ is the solution of the gap equation at $E = B=0$ we find
\begin{equation}
\delta m = \frac{1}{2N_f |E_i(-M_0^2/\Lambda^2)|}
(\Upsilon_1 + \Upsilon_2),
\label{eq:l2}
\end{equation}
where 
\begin{eqnarray}
\Upsilon_1 &=& \frac{q_u^2 + q_d^2}{3M_0^3}{\cal I}_1,\\
\Upsilon_2 &=&- \frac{q_u^4 + q_d^4}{45M_0^7}({\cal I}_1^2 +7 {\cal I}_2^2),
\end{eqnarray}
with ${\cal I}_1 \equiv (eB)^2 - (eE)^2$, ${\cal I}_2 \equiv (eE)(eB)$;
moreover $E_i$ denotes the exponential integral function, $E_i(x) = - \int_{-x}^\infty ds e^{-s}/s$.
The field dependence in the above equation resembles that occurring in the Euler-Heisenberg
lagrangian~\cite{Heisenberg:1935qt} as it should, since the latter can be obtained
by integrating the gap equation over $M_q$.
From Eq.\er{eq:l2} we notice that for $B=0$, $\delta m \propto -E^2/M_0^3$ neglecting
higher order contributions; the curvature of $\delta m$ versus $eE$ does not change as long as
$eE > eB$. For $E  =B$ one has to take into account the contribution $O(E^2B^2)$
which still shows $\delta m \propto -E^2 B^2/M_0^7$ leading to a decreasing $M_q$. 
Finally for $eB>eE$ the catalysis sets in, at least for small values of the fields, 
eventually leading to $\delta m \propto -B^2/M_0^3$ for $E=0$.
In the lower panel of Fig.\ref{Fig:1b} we have compared the perturbative solution
in Eq.\er{eq:l2} with the full one, for two cases. We find a fair agreement among the
two for $eE,eB\simeq 5 m_\pi^2$.

\section{Results at finite temperature}

\begin{figure}[t!]
\begin{center}
\includegraphics[width=7cm]{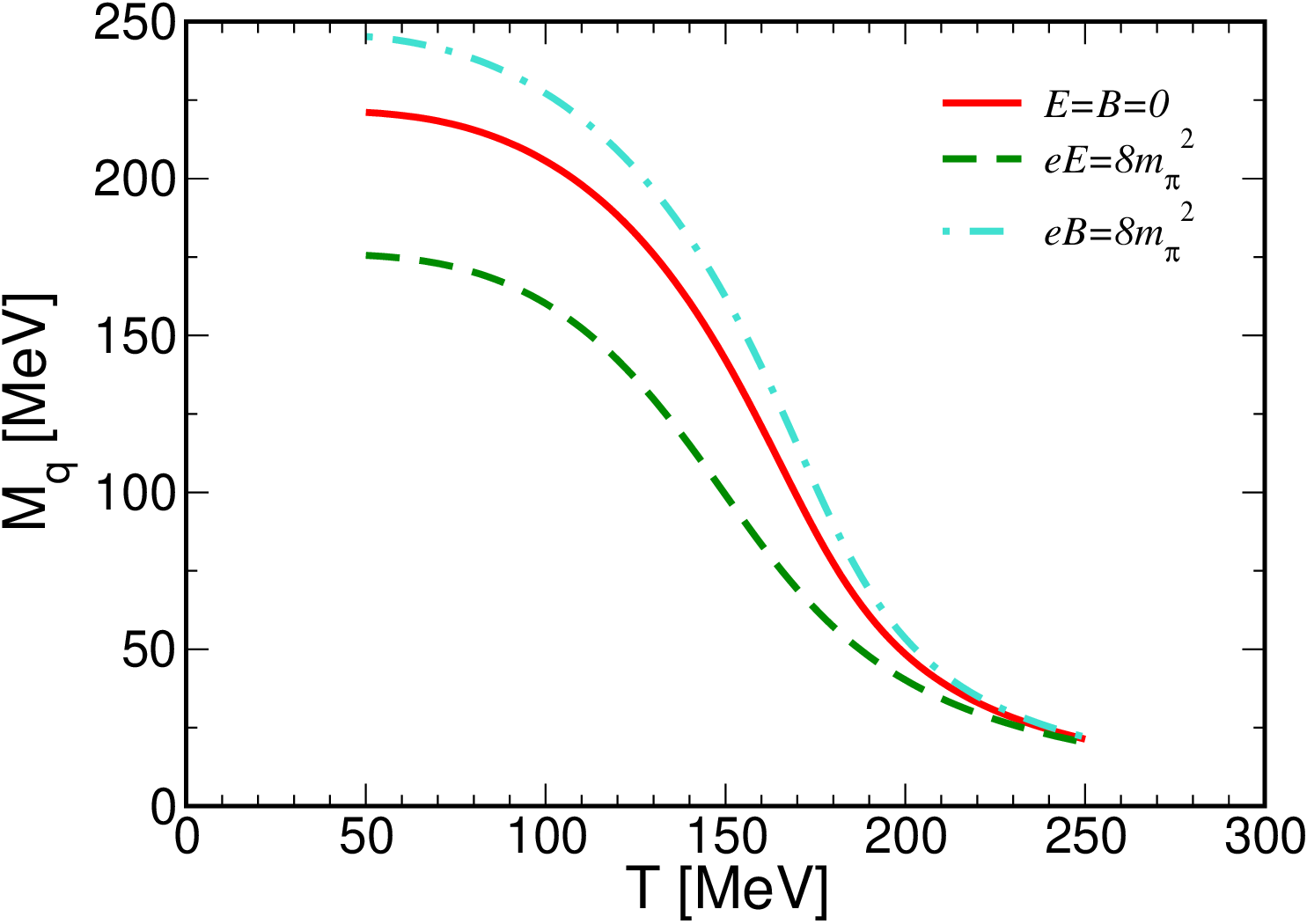}\\
\includegraphics[width=7cm]{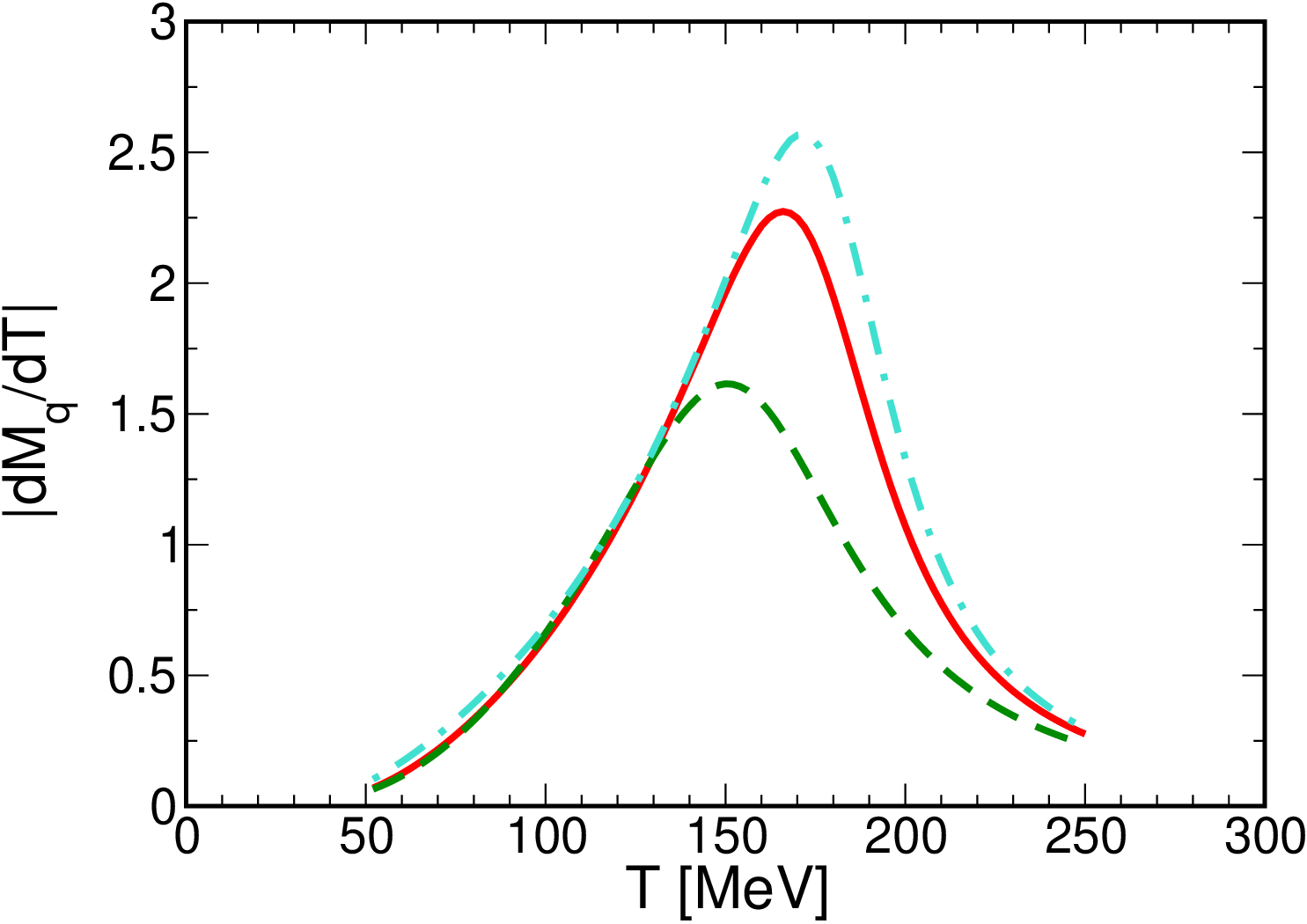}\\
\includegraphics[width=7cm]{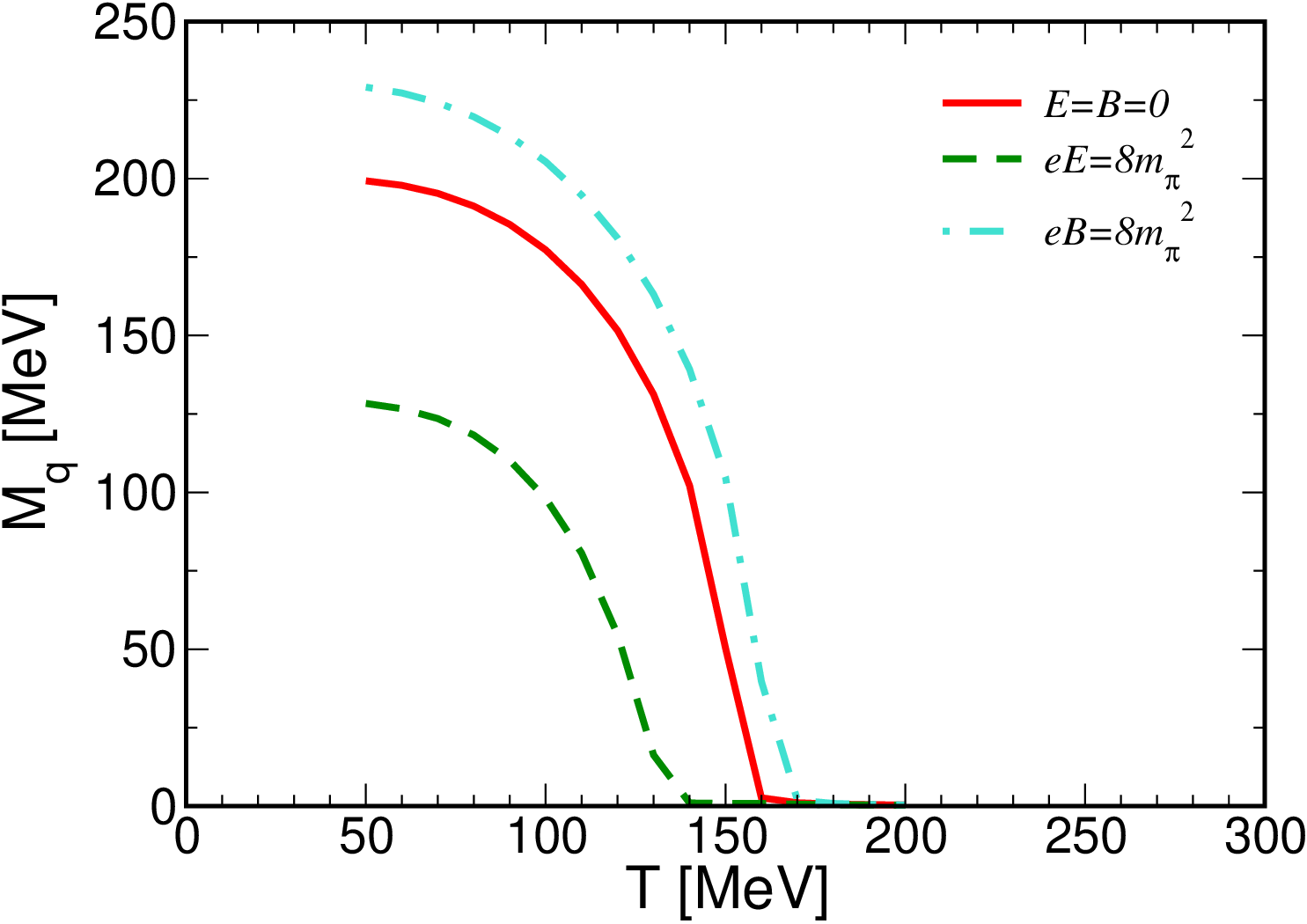}
\end{center}
\caption{\label{Fig:2} {\it(Upper panel)}. Dynamical quark mass versus temperature for
$E=B=0$ (red solid line), $eB =8m_\pi^2$ (cyan dot-dashed line) and $eE=8m_\pi^2$
(dashed green line).
{\it(Middle panel)}. 
$|dM_q/dT|$ versus temperature used to identify the pseudo-critical temperature for the chiral
crossover. Line convention is the same of the upper panel.
 {\it(Lower panel)}. Dynamical quark mass versus temperature for
$E=B=0$ (red solid line), $eB =8m_\pi^2$ (cyan dot-dashed line) and $eE=8m_\pi^2$
(dashed green line) in the chiral limit.}
\end{figure}
In this Section we discuss our results about chiral symmetry restoration at finite 
temperature. In the upper panel of Fig.~\ref{Fig:2} we plot $M_q$  versus $T$ for $E\neq0$ and $B\neq0$,
and compare it with the result at $E=B=0$. The general trend of data shown in the figure 
is in agreement with the scenario depicted at $T=0$ discussed above. In particular the inverse catalysis
due to the electric field implies the lowering of the critical temperature; on the other hand the catalysis
due to the magnetic field at $T=0$ is still present at $T\approx T_c$ leading to the increase of the pseudo-critical
temperature at $B\neq0$. It has been discussed that the magnetic catalysis of chiral symmetry breaking
within the NJL model at finite temperature is due to the fact the NJL interaction kernel 
does not take into account the effects of screening as well as of coupling lowering
which instead occur in QCD and are important for inverse catalysis~\cite{Braun:2014fua,Mueller:2015fka}.
Although it would be possible to insert by hand a $B-$dependence of the NJL coupling
in order to reproduce the inverse magnetic catalysis~\cite{Ahmad:2016iez}, we prefer to not do this 
in our study because it would hide the effect of the electric field; we will add this important
ingredient in our upcoming works.

In the middle panel of Fig.~\ref{Fig:2} we plot $|dM_q/dT|$: we identify its maximum
with the crossover temperature. For $E=B=0$ we find $T_c \approx 166$ MeV. 
We notice that the electric field not only makes the pseudo-critical
temperature lower than the one in the case $E=B=0$, but it also smooths the crossover because the variation
of the quark mass with temperature is smaller in magnitude than in the case with no fields. 
Finally, in the lower panel of Fig.~\ref{Fig:2} we plot $M_q$ versus temperature for the ideal case of
quarks with vanishing current mass: as expected, the effect of the external fields is qualitatively
the same we have found in the realistic case of quarks with finite current mass.

\begin{figure}[t!]
\begin{center}
\includegraphics[width=7cm]{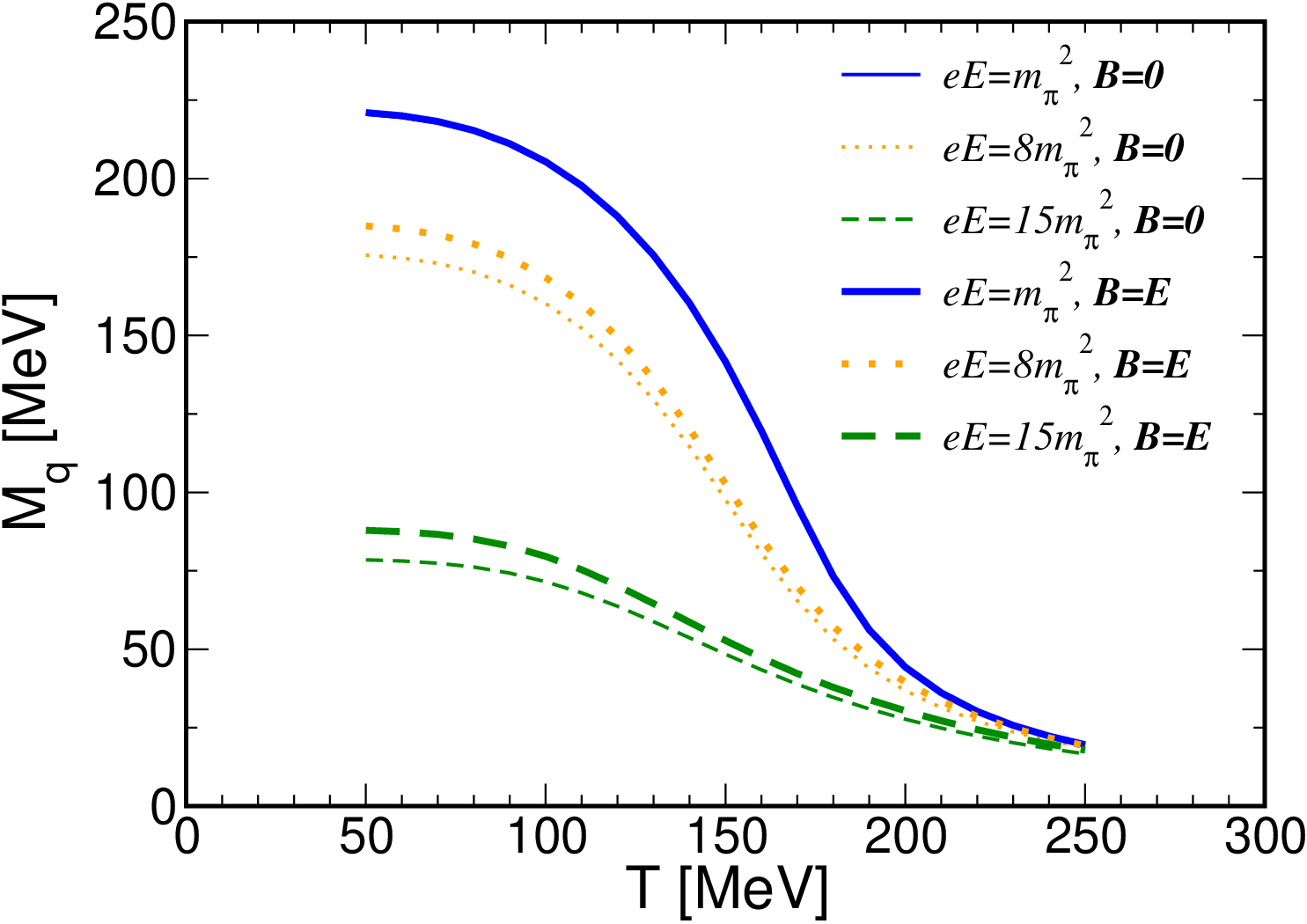}
\end{center}
\caption{\label{Fig:2a} 
Dynamical quark mass versus temperature for several values of $E$ and $B$. 
Thin lines correspond $B=0$, while with thick lines we denote the results for $E = B$.
Blue solid line corresponds to $e E=m_{\pi}^2$, orange dotted line to $eE=8m_{\pi}^2$, and
green dashed line to $eE=15m_{\pi}^2$. Color convention for thick lines 
follows that we have used for thin lines.}
\end{figure}

In Fig.~\ref{Fig:2a} we plot $M_q$ versus $T$ 
for several values of $E$ and $B$: thin lines correspond $B=0$, 
while with thick lines we denote the results for $ E = B$.
Blue solid line corresponds to $eE=m_{\pi}^2$, orange dotted line to $eE=8m_{\pi}^2$, and
green dashed line to $eE=15m_{\pi}^2$. Increasing the electric field strength results in a lowering
of the critical temperature, and the effect of $B\neq0$
is just to increase a bit the quark mass and shift the critical temperature
towards slightly higher values.

The results collected in Figures~\ref{Fig:2} and \ref{Fig:2a} show that even when $B=E$
the effect of the fields on the critical temperature does not cancel and the electric field gives the more
important contribution, leading to an inverse catalysis. In fact one would need a larger value of $B$
to observe an increase of the critical temperature.
This can be understood easily: close to the second order phase transition (we work now at $m_0=0$
which allows an analytical treatment) 
we can make an expansion of the thermodynamic potential in powers of $M_q$, namely
\begin{equation}
\Omega = \frac{\alpha_2}{2}M_q^2+O(M_q^4),
\end{equation}
where the coefficient $\alpha_2 = \partial^2\Omega/\partial M_q^2$ at $M_q=0$;
$\alpha_2$ is negative in the chirally broken phase and vanishes at the phase transition.
The coefficient $\alpha_2$ can be easily computed taking the derivative of the gap equation Eq.\er{eq:gkk}, and expanding
for small values of the fields. It is then possible to write $\alpha_2 = \alpha_{2,0} + \alpha_{2,2}$
where $\alpha_{2,0}$ denotes a field-independent term and $\alpha_{2,2}$ corresponds to a term
$O(eE^2,eB^2)$.
The field-independent term is not interesting because it just determines
the critical temperature when the external fields are set to zero. On the other hand 
the field-dependent contributions are more relevant for the discussion; 
a straightforward calculation leads to
\begin{eqnarray}
\alpha_{2,2}&=&-\sum_f q_f^2\frac{N_c}{4\pi^2}\int ds
~\Theta_3(T,s)
\frac{(eB)^2 - (eE)^2}{3}\nonumber\\
&&-\sum_f  q_f^2\frac{N_c}{48\pi^2T^2}\int \frac{ds}{s}e^{-\frac{1}{4T^2 s}}
\Delta_3(T,s) (eE)^2,\nonumber\\
&&
\label{eq:ops}
\end{eqnarray}
where we have used the shorthand notation 
\begin{eqnarray}
\Theta_3(T,s) &=& \theta_3
\left(
\frac{\pi}{2},e^{-\frac{1}{4T^2 s}}
\right),\\
\Delta_3(T,s) &=&
\left.\frac{d\theta_3(z,x)}{dx}\right|_{z=\frac{\pi}{2},x=\exp[-1/(4T^2 s)]}.
\label{eq:ops2}
\end{eqnarray} 
The term on the right hand side in the first line of Eq.\er{eq:ops} shows that the correction to $\alpha_2$
due to the pure magnetic field is negative, hence shifting the phase transition to larger temperatures.
On the other hand the term proportional to $E^2$ is positive, and gets a further positive contribution
from the second line of Eq.\er{eq:ops}: indeed the latter is proportional to the
derivative of the $\theta_3(x,z)$ function, which is a decreasing function of its second argument.
As a consequence the coefficient proportional to $E^2$ is positive and because of the additional contribution
at finite temperature, one needs a value of $B>E$ in order to change the sign of $\alpha_{2,2}$ and turning
the inverse catalysis into a direct one. This explains why for $E=B$ we still find an inverse catalysis of
chiral symmetry breaking at finite temperature.

\begin{figure}[t!]
\begin{center}
\includegraphics[width=7cm]{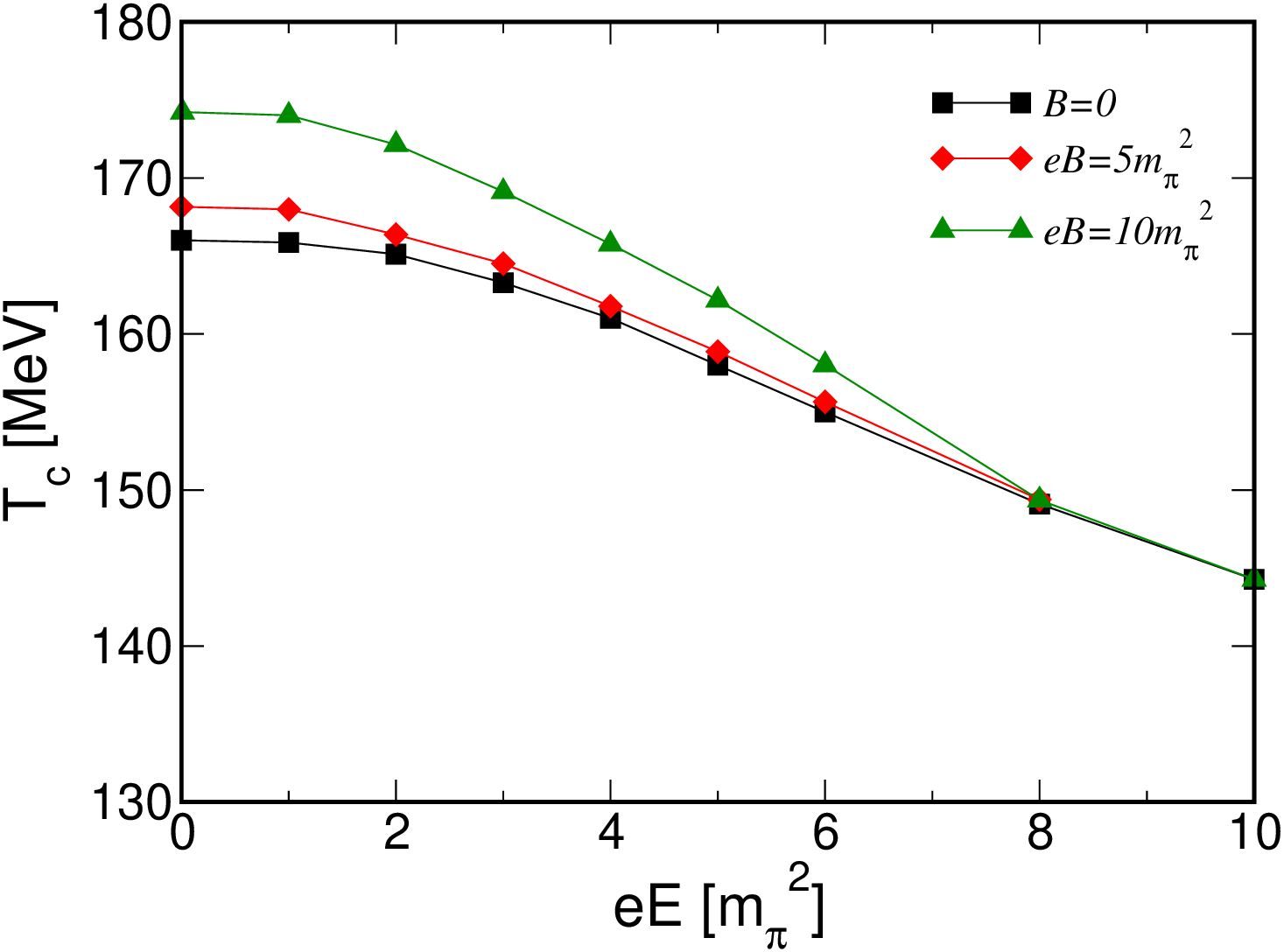}
\end{center}
\caption{\label{Fig:2c} 
Critical temperature for chiral symmetry restoration versus electric field strength, 
measured in units of $m_\pi^2$, for several values of 
the external magnetic field.}
\end{figure}
In Fig.~\ref{Fig:2c} we plot $T_c$ versus $eE$ (measured in units of $m_\pi^2$) 
for several values of the external magnetic field: black squares correspond to $B=0$,
red diamonds to $eB=5 m_\pi^2$ and green triangles to $eB=10 m_\pi^2$. 
This figure summarizes one of the main finding of our work, namely that the electric field
leads to a lowering of the critical temperature for chiral symmetry restoration, and the presence of
the parallel magnetic field does not change this result unless $B\gg E$.

\section{Chiral density effects at the critical temperature\label{eq:cd}}
The electric-magnetic background considered in this article is dynamically unstable
because of the Schwinger pair production~\cite{Heisenberg:1935qt,Schwinger:1951nm}. This is due to the presence
of poles in the thermodynamic potential, $\Omega$, which in turn make the quantum corrections to the
electromagnetic lagrangian complex,
with the imaginary part related to the vacuum persistency probability.
Because of the quantum anomaly, the Schwinger mechanism eventually leads to a nonzero chiral density, $n_5$; 
we assume for the moment the background magnetic field $B$ is very large so that
it is reasonable to assume that only the lowest Landau level (LLL) is occupied, to simplify the discussion;
moreover we assume the system made only of one flavor of quarks, namely $u$ quarks.
Let us focus on one single $u\bar u$ created by the Schwinger effect.
The $u$ quark must have its spin aligned along $\bm B$
because of the LLL approximation; because of dimensional reduction in the LLL $u$  momentum
will be initially rather parallel or antiparallel to $\bm B$, so the initial chirality can be either positive or negative.
On the other hand the effect of $\bm E\parallel \bm B$ is to accelerate 
$u$ along the direction of $\bm B$ so after some time $u$ quark will have positive chirality. An analogous discussion
can be done for the $\bar u$. 
Therefore as a consequence of the Schwinger effect, LLL and $\bm E\parallel \bm B$ each time a pair is created,
there is an increase of a factor two of the net chiral density of the system, $n_5\equiv n_R - n_L$. 
Obviously higher Landau levels do not contribute to $n_5$ because particle spin can be either parallel or antiparallel
to $\bm B$ leading to a cancellation of $n_5$.
Hence chirality is produced dynamically in the background field configuration studied here. 
This makes the study very interesting
because if chiral density relaxes to an equilibrium value, it might affect the equilibrium properties 
of quark matter. 
 
The time evolution of $n_5$ in case of massive particles in the background with constant and homogeneous
fields has been derived for the first time by Warringa in~\cite{Warringa:2012bq}, where he has shown it can be directly obtained from the Schwinger production rate for the case of $\bm E\parallel\bm B$, 
namely~\cite{Fukushima:2010vw,Nikishov:1969tt,Cohen:2008wz,Bunkin:1969if,Dunne:2004nc}
\begin{equation}
\Gamma=\frac{q_f^2 (eE)(eB)}{4\pi^2}\coth\left(\frac{B}{E}\right)
e^{-\frac{\pi M^2}{|q_f eE|}} ;
\end{equation}
indeed only the lowest Landau level (LLL) gives a contribution to $n_5$, and this LLL contribution 
can be easily extracted from the above equation by taking the $B\rightarrow\infty$ limit
because in such a limit it is reasonable to assume that only the LLL is occupied; 
because each pair in the LLL
changes the chiral density of a factor of 2 we have from the above equation in the $B\rightarrow\infty$  limit:
\begin{equation}
\frac{dn_5}{dt}=\frac{q_f^2 (eE)(eB)}{2\pi^2}
e^{-\frac{\pi M^2}{|q_f eE|}} ,
\label{eq:al1}
\end{equation}
in agreement with~\cite{Warringa:2012bq}. If evolution of $n_5$ was given only by the above equation then
the system would never be able to reach thermodynamical equilibrium (assuming the fields as external fields
neglecting any backreaction from the fermion currents). However Eq.\er{eq:al1} is just half of the story:
because of finite quark mass there are chirality changing processes which should lead to equilibration
of $n_5$. In order to take into account of these processes we add a relaxation term on the right hand side
of the above equation,
\begin{equation}
\frac{dn_5}{dt}=\frac{q_f^2 (eE)(eB)}{2\pi^2}
e^{-\frac{\pi M^2}{|q_f eE|}}  - \frac{n_5}{\tau_M},
\label{eq:al2}
\end{equation}
where $\tau_M$ corresponds to the relaxation time of chirality changing processes. 
For $t>>\tau_M$ the solution of Eq.\er{eq:al2} relaxes to teh equilibrium value
\begin{equation}
n_5^{\mathrm{eq}} = \frac{q_f^2 (eE)(eB)}{2\pi^2}
e^{-\frac{\pi M^2}{|q_f eE|}} \tau_M.
\label{eq:al3}
\end{equation}

The equilibrium value of chiral density depends on the value of $\tau_M$. It is reasonable to assume
both by virtue of dimensional considerations and by naive physical arguments that $\tau_M\propto1/M_q$
where $M_q$ corresponds to the constituent quark mass: for large values of $M_q$ chirality changing processes
will be very fast hence reducing drastically the relaxation time and the net chirality produced at equilibrium;
on the other hand for small $M_q$ the system will be less efficient in changing chirality which implies
a larger relaxation time and a larger chirality produced at equilibrium. Thus we assume
\begin{equation}
\tau_M = \frac{c}{M_q};
\label{eq:c}
\end{equation}
the above equation implicitly contains
effects of the chiral condensate in the chirally broken phase via the larger value of $M_q$ in this phase.
Needless to say the parameter $c$ adds the largest uncertainty in our calculations: because $n_5^{\mathrm{eq}}$
depens linearly on $\tau_M$, a change of an order of magnitude in $c$ will produce the same change
in  $n_5^{\mathrm{eq}}$. We will study how changing $c$ might affect our results.

Equation\er{eq:al3} shows that on a time scale larger than the relaxation time 
an equilibrium value of $n_5$, namely $n_5^{\mathrm{eq}}$, is produced. 
Because of the different charges of $u$ and $d$
quarks the equilibrium value of $n_5$ for the two flavors to be different: 
at equilibrium in fact we find
\begin{equation}
\frac{n_{5u}^\mathrm{eq}}{n_{5d}^\mathrm{eq}} =
\frac{q_u^2}{q_d^2} e^{-\frac{\pi M^2}{|eE|}\left(\frac{1}{q_u} - \frac{1}{|q_d|}\right)} ,
\end{equation}  
the actual value depending on $E$ and on temperature via $M_q$.
The existence of an equilibrium value for the chiral density means it is possible to introduce 
a chemical potential for the chiral charge, namely the chiral chemical potential $\mu_5$, 
conjugated to $n_5^{\mathrm{eq}} $.
A self-consistent computation of $\mu_5$ given the value of $n_5^{\mathrm{eq}}$ in Eq.\er{eq:al3}
requires a canonical ensemble calculation in which the gap equation for the quark mass is solved
self-consistently with the number equation, namely 
\begin{equation}
n_5^{\mathrm{eq}} = -\frac{\partial\Omega}{\partial\mu_5},
\label{eq:ne}
\end{equation} 
with $\mu_5$
introduced in the quark propagator with $\bm E \parallel \bm B$.
This full calculation is well beyond the purpose of the present article and is left to a future study.
Here we limit ourselves to consider this problem only in the limit of small
$\mu_5$ as well as small fields,
in which we can use the NJL model with $E=B=0$ but $\mu_5\neq0$
to compute the relation between $\mu_5$ and $n_5^{\mathrm{eq}}$, as well as to take into account 
self-consistently the effect of $\mu_5$ in the gap equation.
The cheap procedure we use here to solve self-consistently the problem should be accurate
up to the lowest nontrivial order in $\mu_5$ and fields, that is $O(\mu_5^2,E^2,B^2)$.

The NJL thermodynamic potential at $E=B=0$ and $\mu_5\neq0$ can be written 
as~\cite{Fukushima:2010fe}
\begin{eqnarray}
\Omega&=&\frac{(M_q-m_0)^2}{4G}  \nonumber\\
&&-N_c\sum_f  T\sum_n\int\frac{d^3\bm p}{(2\pi)^3}
\log(\omega_n^2 + E_+^2)(\omega_n^2 + E_-^2),\nonumber\\
&&
\end{eqnarray}
with $E_\pm^2 = (p\pm\mu_{5f})^2 + M_q^2$ and $\mu_{5f}$ denotes the chiral chemical potential 
for the flavor $f$: we allow for a flavor dependence of $\mu_5$ 
because the equilibrium value of $n_5$ depends on the flavor itself.
At lowest order in $\mu_5$ the correction to the
thermodynamic potential can be written as
\begin{equation}
\delta\Omega = -N_c \sum_f \mu_{5f}^2 T\sum_n\int\frac{d^3\bm p}{(2\pi)^3}
\frac{2(\omega_n^2 +M_q^2 -p^2)}{(p^2 + \omega_n^2 +M_q^2)^2},
\end{equation}
which allows to write the $\mu_5-$dependent correction to the gap equation, namely 
\begin{eqnarray}
\frac{\partial\delta\Omega}{\partial M_q} &=& 
-N_c \sum_f \mu_{5f}^2 T\sum_n\int\frac{d^3\bm p}{(2\pi)^3}
\frac{4M_q(3p^2 -\omega_n^2 -M_q^2)}{(p^2 + \omega_n^2 +M_q^2)^3}.\nonumber\\
&&
\label{eq:pa1}
\end{eqnarray}
Moreover the relation among $n_5=-\partial\Omega/\partial\mu_5$ and $\mu_5$ is given by
\begin{eqnarray}
n_{5f} = \mu_{5f} N_c  T\sum_n\int\frac{d^3\bm p}{(2\pi)^3}
\frac{4(\omega_n^2 -p^2 +M_q^2)}{(p^2 + \omega_n^2 +M_q^2)^2}, \label{eq:pa2}
\end{eqnarray}
and the number equation Eq.\er{eq:ne} can be written as
\begin{equation}
n_{5f} = n_5^{\mathrm{eq}}.
\label{eq:ne2}
\end{equation}
We have verified that in the chiral limit $M_q=0$ the above equation gives
$n_{5f} =  \mu_5 T^2N_c/3 $
in agreement with~\cite{Fukushima:2008xe};
in the case $M_q\neq0$ the relation between $n_5$ and $\mu_5$ is more complicated 
and we have to compute it by  performing numerically the integration in Eq.\er{eq:pa2}.

Taking into account Eq.\er{eq:pa1} the gap equation Eq.\er{eq:gkk} becomes
\begin{eqnarray}
\frac{M_q-m_0}{2G}&=&M_q \frac{N_c}{4\pi^2}\sum_f \int_0^\infty  \frac{ds}{s^2} e^{-M_q^2 s}{\cal F}(s)\nonumber\\
&&+M_q \frac{N_c N_f}{4\pi^2}\int_{1/\Lambda^2}^\infty\frac{ds}{s^2} e^{-M_q^2 s}
-\frac{\partial\delta\Omega}{\partial M_q},\nonumber\\
&&
\label{eq:gem1}
\end{eqnarray}
and $\mu_5$ has to be computed self-consistently according to the number equation, Eq.\er{eq:ne2}.
We notice that although an explicit dependence of $\mu_5$ on $E$ and $B$ is not
present in the above equations, the gap equation Eq.\er{eq:pa2} are coupled
because of the dependence of $n_5^{\mathrm{eq}}$ on the $M_q$.

\begin{figure}[t!]
\begin{center}
\includegraphics[width=7cm]{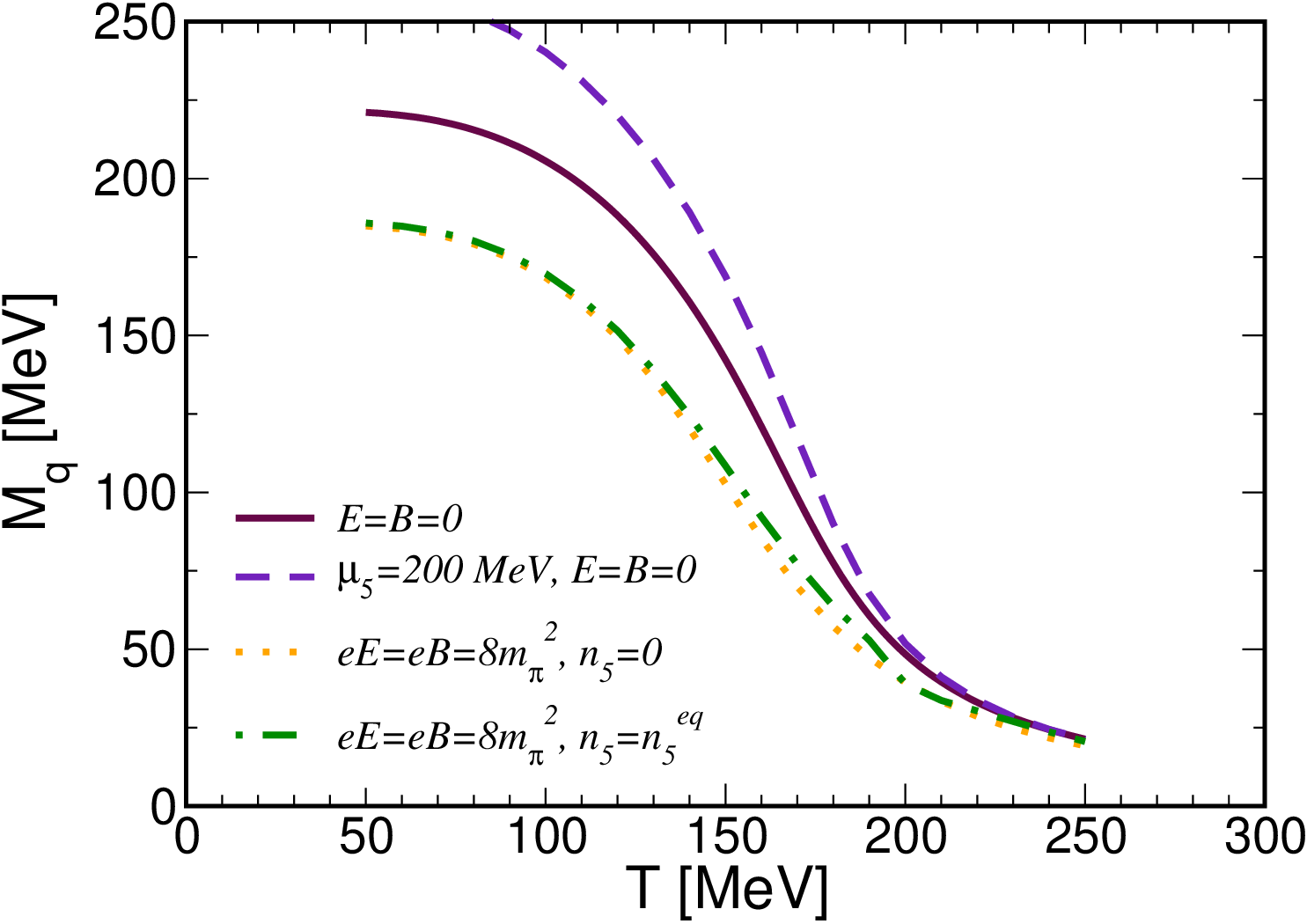}\\
\includegraphics[width=7cm]{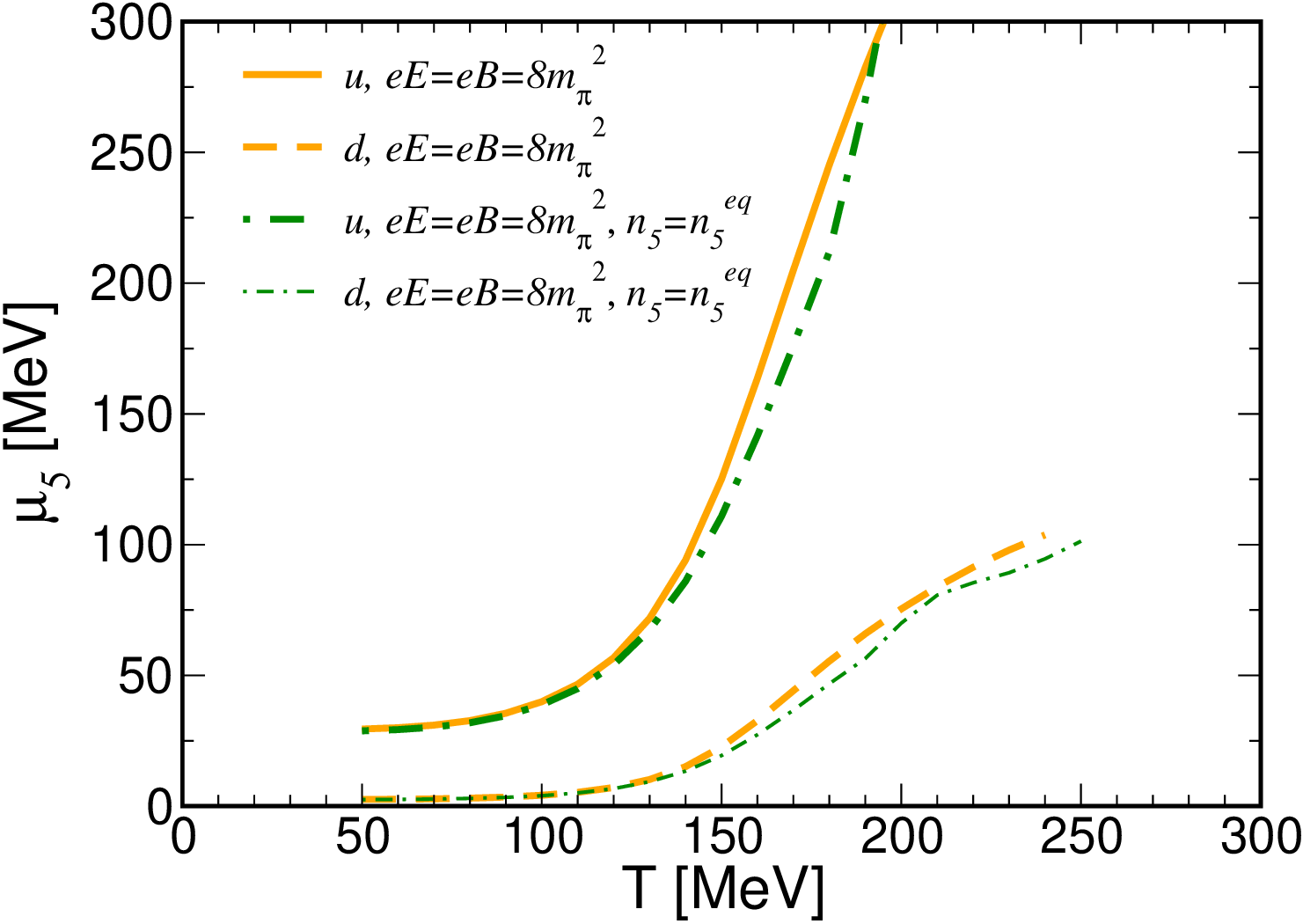}
\end{center}
\caption{\label{Fig:i}  {\it (Upper panel).}  $M_q$ versus temperature. Maroon solid line corresponds to 
$E=B=0$ and $\mu_5=0$. Indigo dashed line corresponds to $E=B=0$ and $\mu_5=200$ MeV
for $u$ and $d$ quarks. 
Orange dotted line corresponds to the case $E=B=8m_\pi^2$ 
and $\mu_5=0$: these data are the same we have shown in Fig.~\ref{Fig:2a}.
Green dot-dashed line corresponds to $E=B=8m_\pi^2$, with both $M_q$ and $\mu_5$ computed
self-consistently by the condition $n_5 = n_{5}^{\mathrm{eq}}$ with $n_5$ given by
Eq.\er{eq:pa2} and the gap equation Eq.\er{eq:pa1}. 
{\it (Lower panel).} Self-consistent $\mu_5$ (green lines) for $u$ (thick dot-dashed line)
and $d$ (thin dot-dashed) quarks versus temperature, corresponding to $M_q$ shown
in the upper panel. For comparison we have also shown
the chiral chemical potential obtained by $M_q$ computed with $\mu_5=0$ shown in the upper panel.}
\end{figure}

In Fig.~\ref{Fig:i} we plot $M_q$ versus temperature for several cases: maroon solid line corresponds to 
$E=B=0$ and $\mu_5=0$. Indigo dashed line corresponds to $E=B=0$ and a common value
$\mu_5=200$ MeV for $u$ and $d$ quarks: we plot
these data to show that the NJL model we use in the calculation is capable to capture the catalysis of chiral
symmetry breaking at finite $\mu_5$ since both $M_q$ and $T_c$ are shifted towards higher values 
in comparison with the case $\mu_5=0$. Orange dotted line corresponds to the case $E=B=8m_\pi^2$ 
and $\mu_5=0$: these data are the same we have shown in Fig.~\ref{Fig:2a}.
Finally green dot-dashed line corresponds to $E=B=8m_\pi^2$, with both $M_q$ and $\mu_5$ computed
self-consistently by solving the number equation Eq.\er{eq:ne2} and the gap equation Eq.\er{eq:pa1}
simultaneously. 
Although $eE=8m_\pi^2$ sounds large for the approximation to be reliable, we present this result
first because it magnifies the effect of the self-consistent $\mu_5$ which would be otherwise
not easy to see. 
In the lower panel of Fig.~\ref{Fig:i}  we plot the chiral chemical potential for 
$u$ (thick dot-dashed line)
and $d$ (thin dot-dashed) quarks versus temperature, corresponding to $M_q$ shown
in the upper panel. For comparison we have also shown
the chiral chemical potential obtained by $M_q$ computed with $\mu_5=0$ shown in the upper panel.
Comparing the results obtained with $\mu_5=0$ and self-consistent $\mu_5$
we notice a slight backreaction of the equilibrium chiral density
on the quark condensate, which reflects in a small change of $M_q$; moreover the catalysis induced by $\mu_5$
is observed thanks to a slight shift of the inflection point of $M_q$ towards larger temperature.
However still the combined effect of $n_5$ at equilibrium and $\bm E \parallel \bm B$ is to 
lower $T_c$ with respect to the case $E=B=0$.

\begin{figure}[t!]
\begin{center}
\includegraphics[width=7cm]{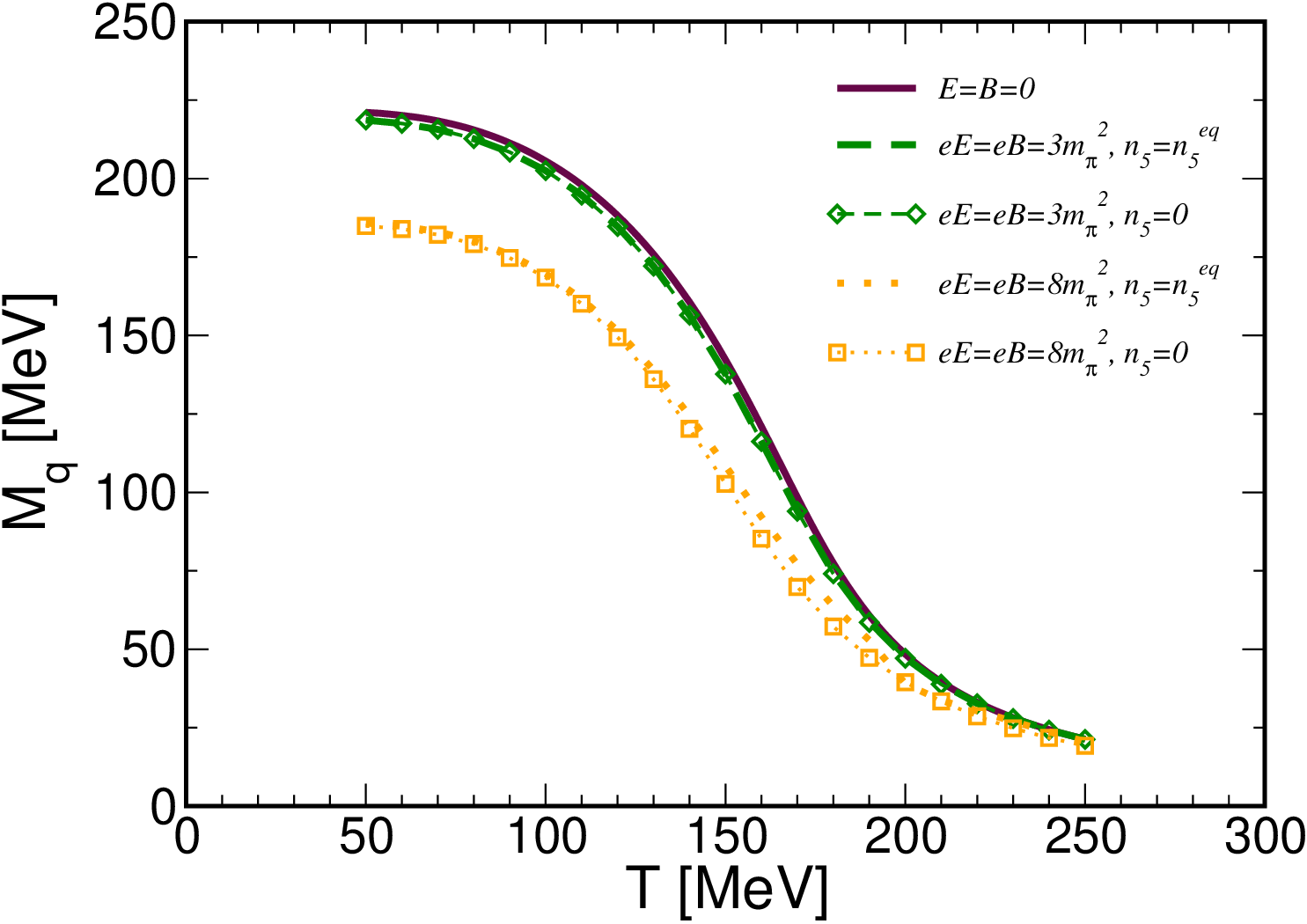}\\
\includegraphics[width=7cm]{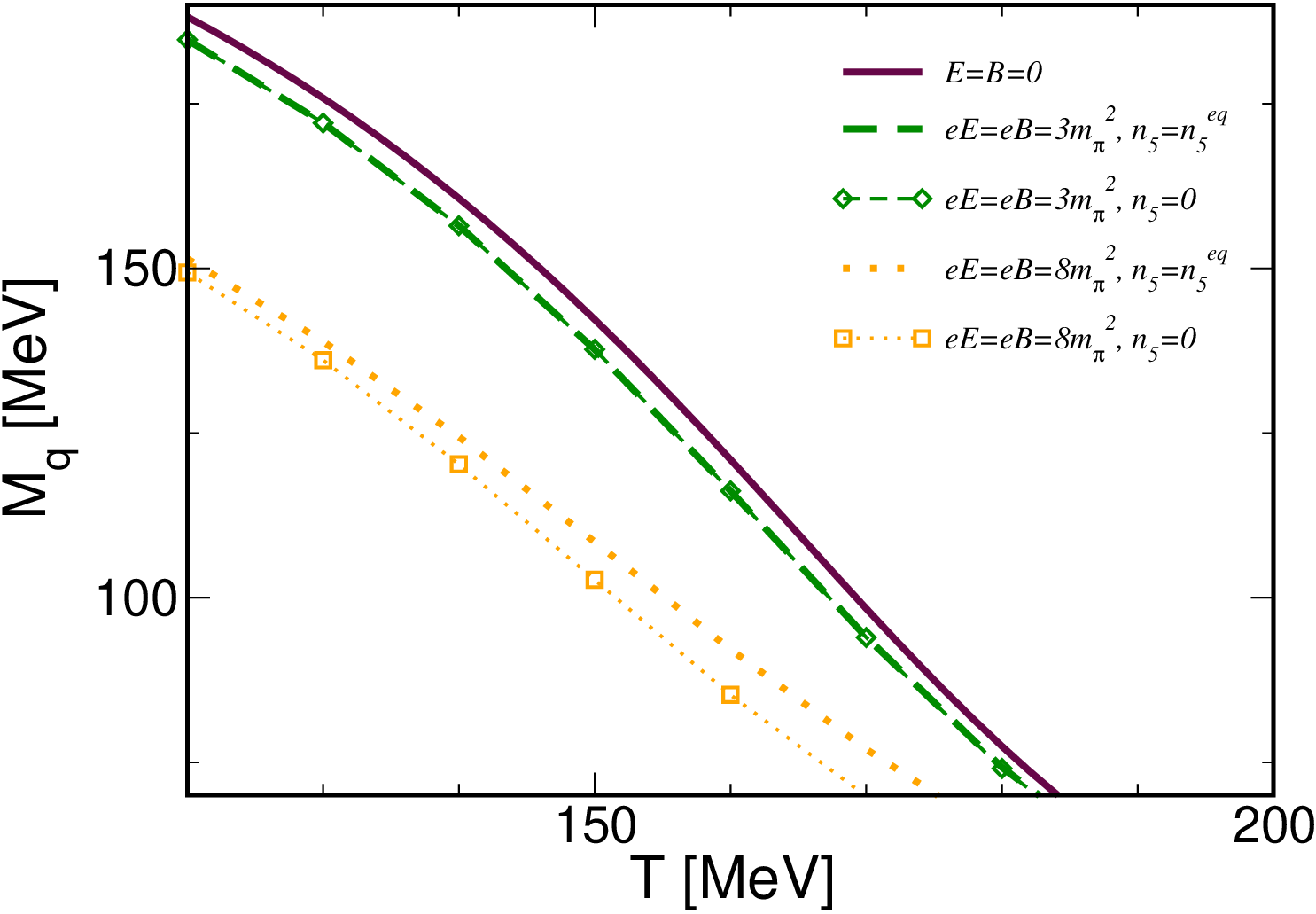}\\
\includegraphics[width=7cm]{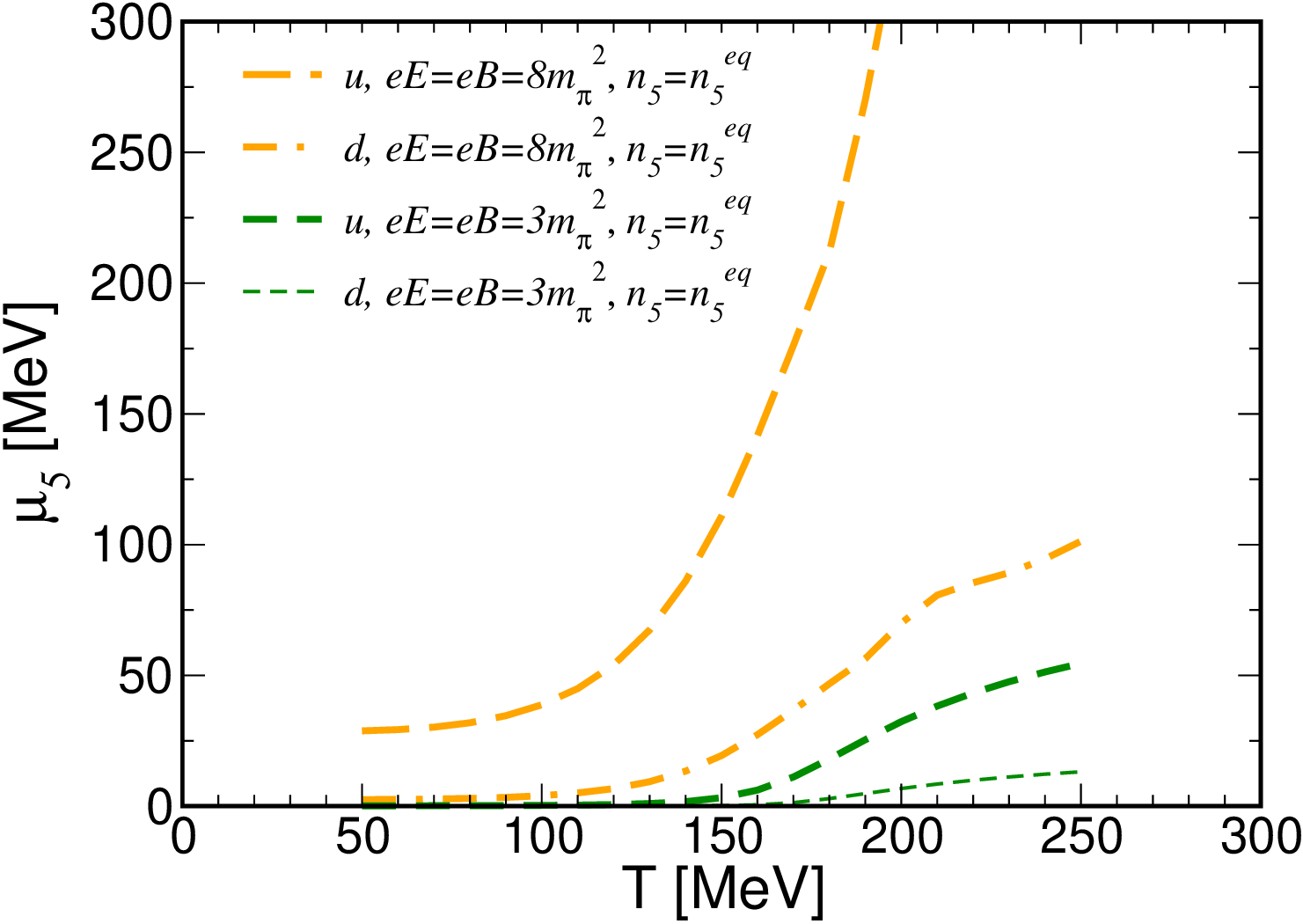}
\end{center}
\caption{\label{Fig:ii}  {\it (Upper panel).}  $M_q$ versus temperature. Maroon solid line corresponds to 
$E=B=0$ and $\mu_5=0$. Indigo dashed line corresponds to $E=B=0$ and $\mu_5=200$ MeV
for $u$ and $d$ quarks. 
Green lines correspond to $E=B=8m_\pi^2$, orange lines to $E=B=8m_\pi^2$. 
Open symbols denote calculations at $\mu_5=0$ while thick lines are for results
with $\mu_5$ computed self-consistently by the number equation and the gap equation.
 {\it (Middle panel)}. Zoom of the upper panel in the temperature range of the chiral crossover.
{\it (Lower panel).} Self-consistent $\mu_5$ for $u$ (thick lines)
and $d$ (thin lines) quarks versus temperature, corresponding to $M_q$ shown
in the upper panel.}
\end{figure}

We have verified that this scenario is in qualitative agreement 
with the one obtained for smaller values of $E$ and $B$, in which case our approximation
should be quantitatively more reliable. In the upper panel Fig.~\ref{Fig:ii} we plot
$M_q$ versus temperature for $E=B=8m_\pi^2$ (orange lines) and
$E=B=3m_\pi^2$ (green lines), as well as for the case $E=B=0$ which we
use as a benchmark (solid maroon line). In the lower panel of the Fig.~\ref{Fig:ii}  we plot the
chiral chemical potentials for $u$ and $d$ quarks at equilibrium computed self-consistently.
We find no qualitative difference between the cases of small and large fields.

For completeness we report the average value of $n_5$ in the crossover region, namely in the temperature range
$(150\div200)$ MeV, which can be obtained directly by using Eq.\er{eq:al3}. We find it runs 
in the range $0.015-0.16$ fm$^{-3}$ in the case of $E=B=3m_\pi^2$,
and  $0.25-1.10$ fm$^{-3}$ in the case of $E=B=8m_\pi^2$. 

The reason why $M_q$  is poorly affected by $\mu_5$ for small values of the fields is 
the different relative change of critical temperature induced by $\mu_5$ on the one hand,
and the electric field on the other hand. 
In the case $eE=eB=3m_\pi^2$ in Fig.~\ref{Fig:ii}, the average values of $\mu_5$ are less than 10 MeV 
in the crossover region. Taking $\mu_5=0$, the effect of $E$ and $B$ is to lower the critical temperature
of about the $5\%$; on the other hand taking $E=B=0$ and $\mu_5=10$ MeV the shift of $T_c$ is practically zero. 
Even increasing by hand the value of mu5 of a factor of 10, the increase of $T_c$ due to mu5 is practically 
negligible compared to the lowering induced by the fields.

\begin{figure}[t!]
\begin{center}
\includegraphics[width=7cm]{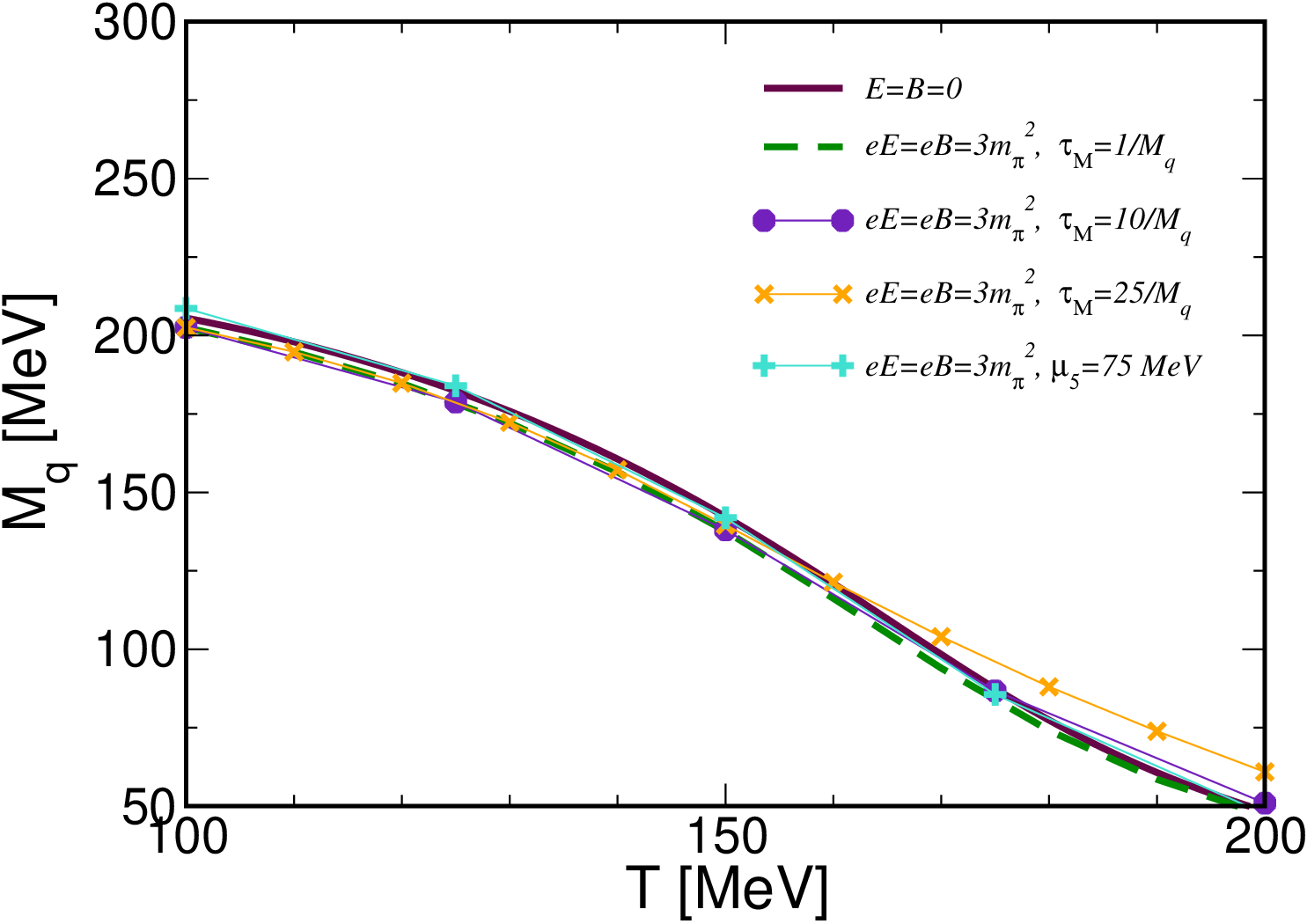}\\
\includegraphics[width=7cm]{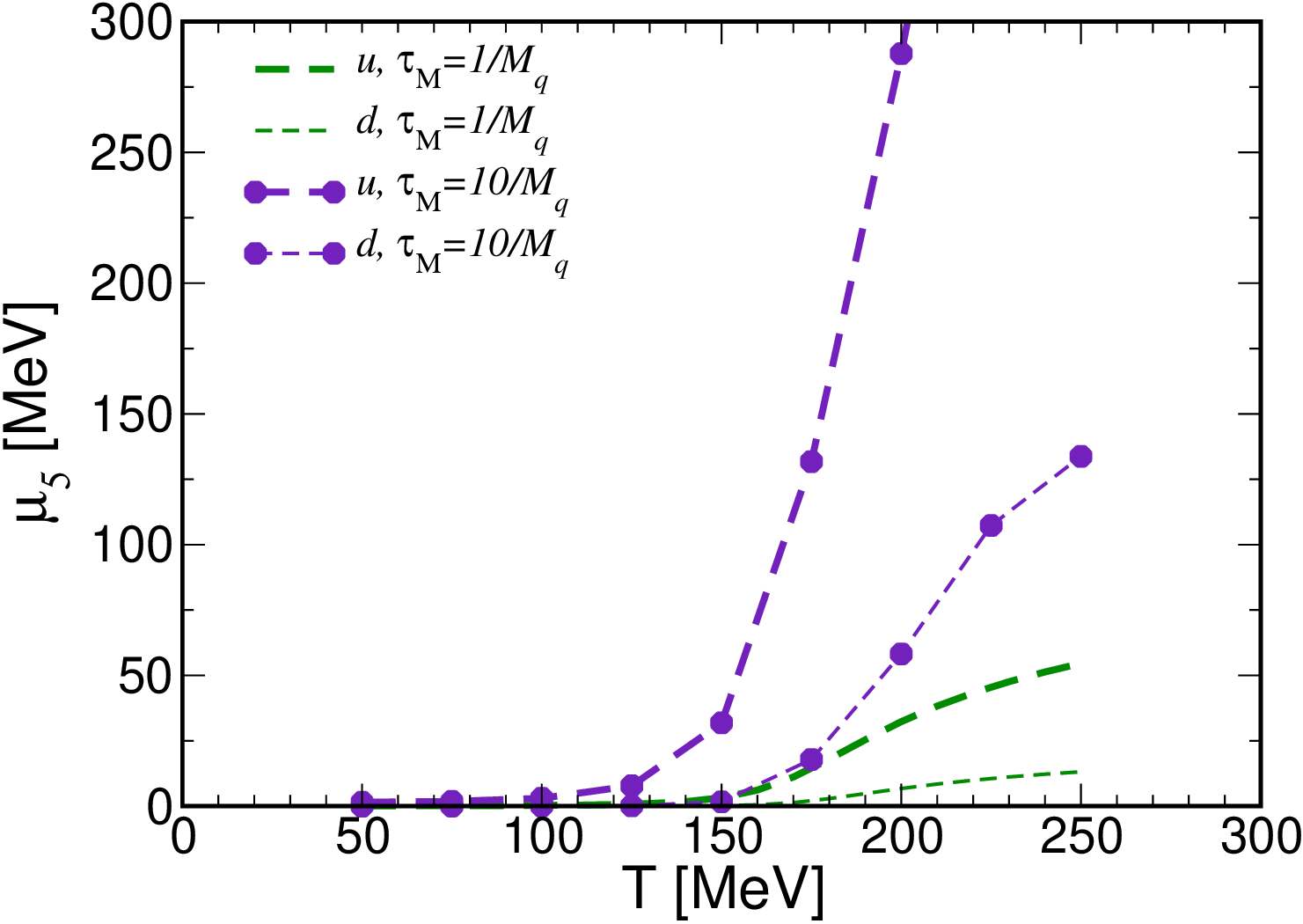}
\end{center}
\caption{\label{Fig:ii2}  {\it (Upper panel).}  $M_q$ versus temperature. Maroon solid line corresponds to 
$E=B=0$ and $\mu_5=0$. 
Green line corresponds to $E=B=3m_\pi^2$ with $\tau_M=1/M_q$. 
Data represented by indigo stars denote $M_q$ computed for $E=B=3m_\pi^2$ with $\tau_M=1/M_q$. 
Orange data correspond to $\tau_M = 25/M_q$.
Finally turquoise plus correspond to a calculation at fixed value of $\mu_5=75$ MeV.
{\it (Lower panel).} Self-consistent $\mu_5$ for $u$ (thick lines)
and $d$ (thin lines) quarks versus temperature, corresponding to $M_q$ shown
in the upper panel.}
\end{figure}

The results shown in Fig.~\ref{Fig:ii} have been obtained for $c=1$ in Eq.\er{eq:c}.
We have checked the stability of the results in the case $eE=3m_\pi^2$
by increasing the relaxation time of an order of magnitude: we collect the results of this check in Fig.~\ref{Fig:ii2}.
In the upper panel of Fig.~\ref{Fig:ii2} we plot $M_q$ versus temperature in the pseudo-critical region.
Maroon and green lines represent the same quantities of Fig.~\ref{Fig:ii}; indigo stars
correspond to $M_q$ computed with $\tau_M=10/M_q$ in Eq.\er{eq:al3}, and 
turquoise plus denote the solution of the gap equation for a fixed value of $\mu_5=75$ MeV. 
We find that for large temperatures the effect of the larger relaxation appears as a tiny shift
of $M_q$ towards larger values; this can be understood because the values of $\mu_{5u}$, $\mu_{5d}$
in this case are larger ot those found with $c=1$, see lower panel of Fig.~\ref{Fig:ii2}.
However still the average value of the chiral chemical potentials is quite small in the pseudo-critical region.
For comparison we have shown the results of a computation at fixed value of $\mu_5=75$ MeV
in the figure: this value of chemical potential 
approximately corresponds to the average value $(\mu_{5u}+\mu_{5d})/2$ computed
self-consistently in the case $\tau_M=10/M_q$ at $T=175$ MeV, see indigo stars in the lower panel of Fig.~\ref{Fig:ii2}.
We find a fair agreement among the calculations with fixed and self-consistent $\mu_5$,
showing that the values of $M_q$ we obtain in the self-consistent calculation are indeed those expected.
We notice that in the case $\tau_M=25/M_q$, shown in Fig.~\ref{Fig:ii2}  by orange data,
we measure a slightly larger increase of $M_q$ due to $\mu_5\neq0$ in the crossover region. 
This result clearly shows how a large $\mu_5$ would affect the thermodynamics balancing the effect
of the external fields; the concrete value of the average $\mu_5$ we have in this case however runs
in the range  $(40\div 320)$ MeV, so the result should not be trusted quantitatively.

\begin{figure}[t!]
\begin{center}
\includegraphics[width=7cm]{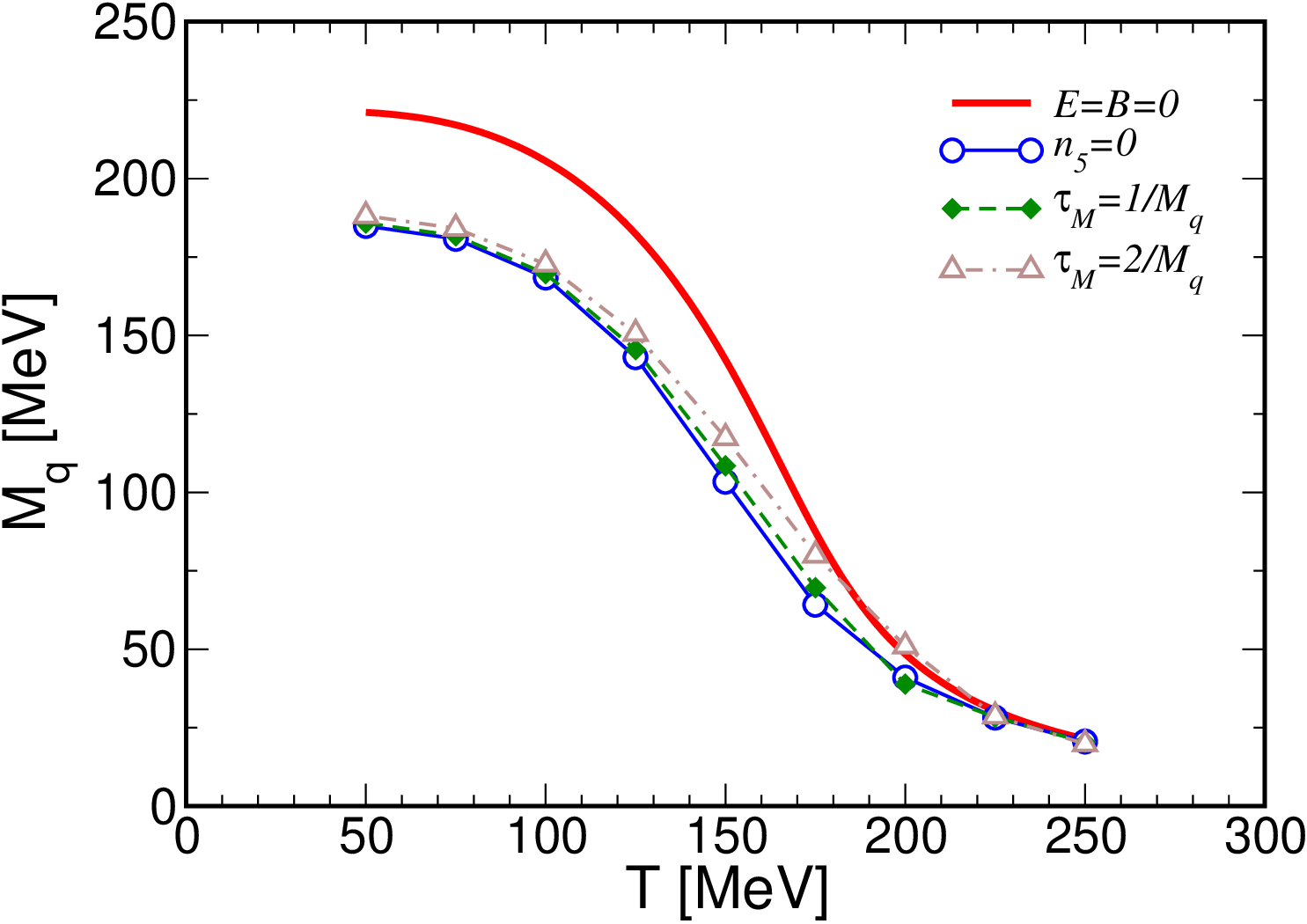}
\end{center}
\caption{\label{Fig:ii2a}  {\it (Upper panel).}  $M_q$ versus temperature for $eE=eB=8 m_\pi^2$. 
As benchmarks we plot by red solid line $M_q$ for the zero field case, and by.
circles $M_q$ with $n_5=0$. 
Diamonds correspond to $\tau_M=1/M_q$, triangles to  $\tau_M=2/M_q$.}
\end{figure}

We have performed the stability check against variations of $c$ or the case $E=B=8m_\pi^2$ discussed above.
We plot in Fig.\ref{Fig:ii2a} the result of this check for the cases of $\tau_M=1/M_q$ (diamonds)
and $\tau_M = 2/M_q$ (triangles).
We have found that taking $c=2$ affects $M_q$ considerably, 
hence showing a net effect of chiral density on the phase transition.
However we take this result not too seriously because doubling
$c$ would roughly correspond to double $\mu_5$, which is already quite large in the pseudo-critical region
as it is shown in Fig.~\ref{Fig:ii} hence making the use of our approximation questionable.
Our conclusion is that as long as the values of $E$ and $\mu_5$ are not too large, our approximate
solution to the self-consistent problem is fairly good, while for larger values of the background field
it has to be taken with a grain of salt.

\begin{figure}[t!]
\begin{center}
\includegraphics[width=7cm]{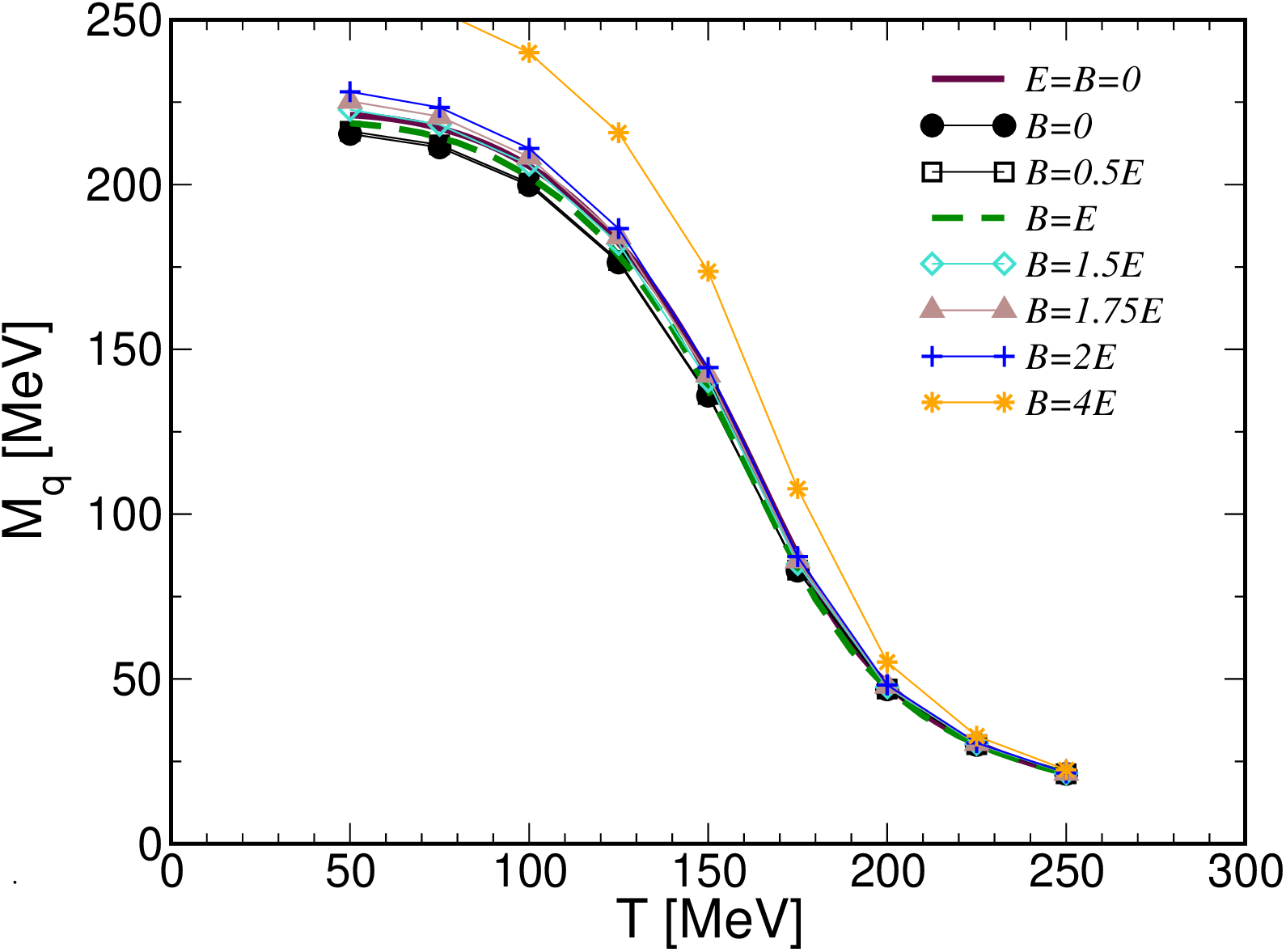}
\end{center}
\caption{\label{Fig:iiab} $M_q$ versus temperature for $eE=3m_\pi^2$ and several values
of $eB$.}
\end{figure}

In Fig.~\ref{Fig:iiab} we plot $M_q$ versus temperature for $eE=3m_\pi^2$ and several values of $eB$.
The computations have been performed by taking into account the dynamically generated $n_5$
for $u$ and $d$ quarks. It is interesting that even if the magnetic field acts as a catalyzer of chiral symmetry breaking
at small temperatures, the presence of the electric field helps an inverse catalysis in the critical region: 
for example in the case $B=2E$ we find that $M_q$ close to the critical temperature
still sits on the zero field result; increasing the magnitude of $E$ just moves the critical temperature
to a lower value.  This is in agreement with the analytical discussion of Section IV.

\section{Conclusions}
In this article we have studied spontaneous chiral symmetry breaking for quark matter
in the background of static, homogeneous 
and parallel electric field $\bm E$ and magnetic field $\bm B$.
We have used a Nambu-Jona-Lasinio model with a local kernel interaction to compute the relevant quantities
to describe chiral symmetry breaking at finite temperature for a wide range
of $E$ and $B$.

Part of our study has been devoted to a mean field calculation of the response of the chiral condensate
to the external fields, both at zero and at nonzero temperature. We have derived both numerically and analytically
the magnetic catalysis and the electric inverse catalysis at zero temperature; we have also
studied the behaviour of the quark condensate at finite temperature, finding a competition between the 
magnetic and electric fields which affects the critical temperature. We have not considered a by-hand modification
of the NJL coupling constant in order to reproduce inverse magnetic catalysis for small $B$ at finite temperature,
because this would have masked the genuine response of the model to an electric field, but we will certainly
consider this necessary modification to the interation term in the future. Our result in this direction is that
the critical temperature for chiral symmetry restoration, $T_c$, is lowered by the simultaneous presence
of the parallel electric and magnetic fields.

We have then focused on
the effect of equilibration of chiral density, $n_5$, produced dynamically by axial anomaly on the critical temperature.
Chiral density is produced thanks to Schwinger tunneling and spin alignment in the magnetic field.
The equilibration of $n_5$ happens as a consequence of chirality flipping processes in the thermal bath; 
we have introduced the relaxation time for chirality, namely $\tau_M$, giving the time scale
necessary for the equilibration of $n_5$. 
In absence of a specific calculation of $\tau_M$ it is possible to give only an ansatz;
we chose $\tau_M \propto 1/M_q$ where $M_q$ is the constituent quark mass. 

Because this dynamical system reaches a thermodynamical equilibrium state
for $t\gg\tau_M$, with a specified value of $n_5=n_5^{\mathrm{eq}}$ depending on the actual values of the field and of the temperature,
it is possible to introduce the chiral chemical potential, $\mu_5$, conjugated to $n_5^{\mathrm{eq}}$ at equilibrium.
The value of $\mu_5$ has been computed by coupling the gap equation to the number equation, 
at the leading order in $eE/T^2$, $eB/T^2$ and $\mu_5/T$.
Because of the different electric charges of $u$ and $d$ quarks at equilibrium 
$n_{5u}^{\mathrm{eq}} \neq n_{5d}^{\mathrm{eq}}$ and the ratio of the two is about $5\div6$ 
in the critical region; we have therefore introduced two chemical potentials,
$\mu_{5u}$ and $\mu_{5d}$ conjugated respectively to $n_{5u}^{\mathrm{eq}}$ and $n_{5d}^{\mathrm{eq}}$.

We have found that the equilibrated chiral density 
does not change drastically the thermodynamics
as long as $\mu_5$ at equilibrium is not too large; namely, the inverse catalysis effect
induced by the background fields is not spoiled by the presence of the $\mu_5$ background.
The weak effect of $\mu_5$ on the shift of $T_c$ in presence of the background fields
can be understood because the change of $T_c$ induced by $\mu_5$ itself are smaller than the ones
induced by the background fields.
For example in the case $\mu_5=0$, the effect of the background fields is to lower the critical temperature
of about the $5\%$; on the other hand taking $E=B=0$ and $\mu_5=10$ MeV
which corresponds to the average value of the chemical potential we find in the crossover
region, the shift of $T_c$ is practically zero. 
This conclusion might be no longer valid in the case of large $\mu_5$. For larger values of the fields
we have found that $M_q$ is effectively pushed towards larger values in the critical region by $\mu_5\neq0$.
A firm conclusion about this finding can be achieved however only
by solving the problem beyond the perturbative analysis used in our study.

We would like to remark that the results presented here have to be
considered only explorative: the study of this problem beyond the the weak field and small $\mu_5$ 
approximation will be the topic of upcoming research.
Moreover, the theoretical calculation of the equilibrium value of $\mu_5$
has an uncertainty because of the lack of information about the relaxation time for chirality
flipping processes, $\tau_M$ in Eq.\er{eq:c}, whose computation will be the theme of near future research.

{\em Acknowledgments}. The authors would like to thank the 
CAS President's International Fellowship Initiative (Grant No. 2015PM008), 
and the NSFC projects (11135011 and 11575190).

\end{document}